\documentclass[fleqn,usenatbib]{mnras}
\usepackage{newtxtext,newtxmath}
\usepackage[T1]{fontenc}
\usepackage{amsmath,amssymb,amsfonts}
\usepackage{graphicx}
\usepackage{booktabs}
\usepackage{hyperref}
\usepackage{siunitx}
\usepackage{orcidlink}
\usepackage{xcolor}

\newcommand{\Am}[1]{A_{#1}}
\newcommand{\Msun}{\mathrm{M}_{\odot}}
\newcommand{\kpc}{\,\mathrm{kpc}}
\newcommand{\kms}{\,\mathrm{km\,s^{-1}}}
\newcommand{\Rd}{R_{\rm d}}
\newcommand{\Nboot}{N_{\rm boot}}
\newcommand{\Ndisk}{N_{\rm disk}}
\newcommand{\Amone}{\ensuremath{\Am{1}}}
\newcommand{\Amtwo}{\ensuremath{\Am{2}}}
\newcommand{\pipeline}{\textsc{TNGGalaxyLab}}

\DeclareRobustCommand{\VAN}[3]{#2}
\let\VANthebibliography\thebibliography
\def\thebibliography{\DeclareRobustCommand{\VAN}[3]{##3}\VANthebibliography}

\title[Lopsidedness in TNG50 and TNG100 discs]{Shot noise, aperture and resolution in Fourier measurements
  of disc lopsidedness: TNG50-1 and TNG100-1 analysed with
  \pipeline}

\author[A. U. Omonov et al.]{%
    A. U. Omonov$^{1}$\thanks{E-mail: abbosomonov998@gmail.com}\orcidlink{0009-0006-7897-1255}, K. T. Mirtadjieva$^{1,2}$\thanks{E-mail: mkt1959@mail.ru}\orcidlink{0000-0002-1172-9054}, S. N. Nuritdinov$^{1}$\thanks{E-mail: nur200848@mail.ru}, D. Dottorini$^{3,4}$\thanks{E-mail: dario.dottorini@inaf.it}\orcidlink{0009-0004-5511-3309}\\
  $^{1}$ National University of Uzbekistan, Tashkent, 100174, Uzbekistan.\\
   $^{2}$ Astronomical Institute of the Academy of Sciences of the Republic of Uzbekistan, Tashkent, 100052, Uzbekistan.\\
   $^{3}$ INAF - Osservatorio Astronomico di Roma, Via Frascati 33,
  00078 Monte Porzio Catone, Italy.\\
   $^{4}$ Dipartimento di Fisica, Universit\`a di Roma Tor Vergata,
  Via della Ricerca Scientifica 1, 00133 Roma, Italy.
}

\date{Accepted XXX. Received YYY; in original form ZZZ}

\pubyear{\the\year{}}
\usepackage{tikz}
\usetikzlibrary{shapes.geometric, arrows.meta, positioning, fit, backgrounds}
\usepackage[ruled,linesnumbered,vlined]{algorithm2e}
\begin{document}
\label{firstpage}
\pagerange{\pageref{firstpage}--\pageref{lastpage}}
\maketitle

\begin{abstract}
We present \pipeline, an open-source Python pipeline for particle-based Fourier decomposition of the stellar surface density of simulated disc galaxies.  It measures $m=1$--4 amplitude and phase profiles in the \citet{1995ApJ...447...82R} normalisation, handles IllustrisTNG snapshot conventions, and recovers injected amplitudes in synthetic discs to better than 2 per cent at $N = 5\times10^5$. Fourier amplitudes are positive-definite, so Poisson noise biases them upwards by an amount that grows as the particle number falls. We measure a per-galaxy noise floor by randomising particle azimuths and define a noise-corrected amplitude $A_{1,\rm corr}$.  Downsampling TNG50-1 discs to the TNG100-1 particle mass inflates the uncorrected $\Amone$ by 40--60 per cent, while $A_{1,\rm corr}$ is recovered to within $\sim10$ per cent.  We then process all central galaxies with $3.16\times10^9 < M_\star/\Msun < 10^{11}$ in TNG50-1 (915) and TNG100-1 (6717) with identical settings in the $[1.5,2.5]\,R_{\rm d}$ aperture.  Removing the bootstrap and shot-noise biases lowers the TNG100/TNG50 ratio of median $\Amone$ from 2.3 to 1.6.  At fixed stellar mass, kinematically selected TNG100-1 discs remain 1.5--1.9 times more lopsided after correction; this excess persists whether the aperture is defined in $R_{\rm d}$ or in kpc, is not explained by $\kappa_{\rm rot}$ or $D/T$, and is largest at 3.5--6\,kpc.  TNG50-1 is statistically consistent with the mass-matched S$^{4}$G sample of \citet{2013ApJ...772..135Z}; TNG100-1 is not.  The low $\Amone$ reported for TNG100 by \citet{2022A&A...662A..53L} cannot be read as a resolution deficit, since with matched measurements TNG100-1 is more lopsided than TNG50-1. Code and catalogues are public.
\end{abstract}




\section{Introduction}
\label{sec:intro}

\subsection{The Fourier decomposition of disc surface density}
Lopsidedness - a global $m=1$ asymmetry of the stellar or gaseous disc - has been recognised as a common feature of disc galaxies since the early H\,\textsc{i} studies of \citet{1980MNRAS.193..313B} and
\citet{1994A&A...290L...9R}, who found that up to half of all spirals show significant asymmetry.  Subsequent H\,\textsc{i} profile surveys~\citep{1998AJ....115...62H,1998AJ....116.1169M} and group-environment studies \citep{2006MNRAS.369.1849A} confirmed this ubiquity, while optical and near-infrared photometric surveys quantified the stellar lopsidedness distribution in the local Universe \citep{1995ApJ...447...82R,1997ApJ...477..118Z,2008ApJ...677..186R,2009ApJ...691.1005R,2013ApJ...772..135Z}.
Kinematic lopsidedness has likewise been mapped in resolved H\,\textsc{i} velocity fields \citep{2011A&A...530A..29V}.

Several physical mechanisms have been proposed for the origin of disc lopsidedness: asymmetric cold-gas accretion \citep{1997ApJ...488..642J,2005A&A...438..507B,2008A&ARv..15..189S}, minor mergers and
tidal interactions \citep{1995ApJ...448...41H,1996ApJ...460..121W}, the response of the disc to a perturbed or offset dark-matter halo \citep{1994ApJ...420..597W,1998MNRAS.299..499W,1998ApJ...496L..13L,2001MNRAS.328.1064N,2002ApJ...568..190I},
and long-lived global $m=1$ instabilities of the disc itself \citep{2007MNRAS.382..419S,2008MNRAS.387....2D}. For a comprehensive review, see \citet{2009PhR...471...75J}.
Non-axisymmetric structures in galactic discs - bars, spiral arms, and lopsidedness - carry information about the dynamical state, merger history, and dark-matter halo properties of individual galaxies~\citep{2008gady.book.....B,1993RPPh...56..173S}.  The most widely used quantitative tool for describing such structure is the Fourier decomposition of the projected stellar surface density, $\Sigma(R,\phi)$, into modes of integer azimuthal wavenumber $m$ \citep[R\&Z95 hereafter]{1995ApJ...447...82R}:
\begin{equation}
  \Sigma(R,\phi) = \Sigma_0(R) \left[ 1 +
      \sum_{m=1}^{\infty} \Am{m}(R)\,\cos\!\bigl(m[\phi - \phi_m(R)]\bigr)
  \right],
  \label{eq:fourier}
\end{equation}
where $\Am{m}(R)$ is the fractional amplitude of the mode and $\phi_m(R)$ its phase.  The $m=1$ (lopsided) and $m=2$ (bar/spiral)
modes have received extensive attention because they trace, respectively, the response of the disc to external perturbations
\citep{2002A&A...391..471J,2005A&A...438..507B} and internal instabilities \citep{2003MNRAS.341.1179A,2007MNRAS.382..419S}.  Fourier decomposition complements non-parametric asymmetry measures such as the rotational-asymmetry index \citep{1996ApJS..107....1A,2003ApJS..147....1C} and other non-parametric morphological statistics such as Gini-$M_{20}$~\citep{2004AJ....128..163L}, with the advantage of separating individual azimuthal modes ($m = 1$ lopsidedness vs.\ $m = 2$ bars/spirals) and providing radial profiles of each.

\subsection{Motivation for a reproducible pipeline}

Despite the ubiquity of Fourier analysis in extragalactic studies, published implementations of the method vary significantly in normalisation conventions, radial binning choices, and error estimation procedures.  A factor-of-two ambiguity exists between the
R\&Z95 normalisation and the alternative used by \citet{2007MNRAS.382..419S}, and different authors adopt distinct definitions
of the ``bar length'' \citep[cf.][]{2003MNRAS.341.1179A,2005A&A...434..109A, 2009A&A...495..491A}.  Reported uncertainties often lack a clear description of the underlying assumptions.  Small differences in the adopted disc orientation, centring, or aperture can shift $\Amone$ by $>10\%$, larger than the statistical error in modern cosmological simulations.

The rise of large simulation suites such as EAGLE \citep{2015MNRAS.446..521S}, IllustrisTNG \citep{2019ComAC...6....2N} and
FIRE-2 \citep{2018MNRAS.480..800H} makes reproducibility especially important: catalogues of Fourier amplitudes are increasingly used as inputs to secondary studies of galaxy formation, halo assembly, and observational biases.  Recent work by \citet{2022A&A...662A..53L} systematically analysed 1912 disc galaxies from IllustrisTNG100 and found a lopsided fraction ($\Amone > 0.1$) of $\sim$8\%, considerably lower than the $\sim$30\% inferred from observations \citep{2009PhR...471...75J}, attributed provisionally to resolution limitations of TNG100 and to the ``overquenching'' tendency of the IllustrisTNG feedback model.  In parallel, \citet{2023MNRAS.523.5853V}, \citet{2023MNRAS.526.1636D}, \citet{2025A&A...693A..63D} and \citet{2025A&A...693A..64F} have built large TNG50 lopsidedness catalogues to study the physical drivers of disc asymmetry, its redshift evolution and its machine-learning classification.  \citet{2023MNRAS.526.1636D} also showed that the lopsided fraction in TNG50 changes substantially with the radial range over which $\Amone$ is averaged.  Numbers obtained with different apertures, normalisations and sample definitions are therefore not directly comparable, and a difference between two studies says little about the simulations themselves unless the measurement is held fixed.

A second, less discussed issue is statistical.  The amplitude $\Amone$ is the modulus of a complex sum and cannot be negative.
Poisson fluctuations in the particle distribution therefore add a positive bias, of order $\sqrt{\pi/N}$ for $N$ particles in an
annulus, and this bias is larger in lower-resolution runs.  A resampling bootstrap applied on top of a noisy estimate adds a
further, similar bias.  Neither effect matters much for a well-sampled TNG50 disc, but both become important when amplitudes
from simulations with very different particle masses are compared.

General-purpose simulation-analysis frameworks such as \textsc{PYNBODY} \citep{2013ascl.soft05002P}, \textsc{YT} \citep{2011ApJS..192....9T}, and the \textsc{ILLUSTRIS\_PYTHON} module \citep{2019ComAC...6....2N} provide efficient particle
I/O and generic reduction primitives. Still, none of them includes a validated, error-quantified Fourier decomposition tailored to lopsidedness and bar analysis.  \pipeline, the pipeline presented in this paper, complements these frameworks in three ways: (i) a synthetic-galaxy validation suite that quantifies the recovery bias and shot-noise scaling of each mode (Section \ref{sec:validation}); (ii) bootstrap uncertainties together with an explicit, per-galaxy estimate of the shot-noise floor, so that amplitudes measured at different resolutions can be compared (Section \ref{sec:noise_floor}); and (iii) consistent handling of IllustrisTNG snapshot conventions (the $\sqrt{a}$ velocity factor and the periodic box), which generic loaders do not apply automatically.

\subsection{This work}

We present \pipeline, a Python package for particle-based Fourier analysis of simulated disc galaxies. Our specific contributions
are:
\begin{enumerate}
  \item An implementation of Eq.\,\ref{eq:fourier} in the R\&Z95 normalisation for direct particle sums;
  \item A phase-coherent bar-length estimator in the spirit of \citet{2005A&A...434..109A} and \citet{2009A&A...495..491A};
  \item An optional $\Sigma(R)$-weighted global average in the integral form of \citet{2002A&A...391..471J};
  \item Shrinking-sphere centring \citep{2003MNRAS.338...14P}, the $\sqrt{a}$ velocity conversion \citep{2005MNRAS.364.1105S,2019ComAC...6....2N} and periodic-boundary unwrapping for cosmological snapshots;
  \item Bootstrap uncertainties with $\Nboot = 200$, a per-galaxy shot-noise floor and a noise-corrected amplitude $A_{1,\rm corr}$, validated by downsampling TNG50-1 to the TNG100-1 particle mass;
  \item Validation against synthetic galaxies with analytic amplitudes;
  \item Application with identical settings to 915 TNG50-1 and 6717 TNG100-1 central galaxies, a decomposition of the difference between the two runs into measurement biases and a residual that is not explained by the measurement, and a comparison with the S$^{4}$G sample of \citet{2013ApJ...772..135Z} and with \citet{2022A&A...662A..53L}.
\end{enumerate}

Section \ref{sec:algorithm} describes the algorithm and Section~\ref{sec:validation} the synthetic-galaxy tests. Section \ref{sec:tng50} presents the TNG50-1 sample, Section \ref{sec:comparison} compares TNG50-1, TNG100-1 and the observations and separates measurement bias from resolution dependence, Section \ref{sec:systematics} discusses systematic uncertainties, and Section \ref{sec:astrophysics} the astrophysical implications of our analysis.  Finally, Section \ref{sec:conclusions} summarises our results.

\section{Algorithm}
\label{sec:algorithm}

\begin{figure*}
  \centering
  \begin{tikzpicture}[node distance=0.9cm and 0.7cm, every node/.style={font=\small},
      inputbox/.style={rectangle, rounded corners, draw=black!70!black, fill=white!10, thick, minimum width=3cm, minimum height=1cm, align=center},
      processbox/.style={rectangle, rounded corners, draw=black!80, fill=white!10, thick, minimum width=3.4cm, minimum height=1.1cm, align=center},
      decisionbox/.style={rectangle, rounded corners, draw=black!80, fill=white!15, thick, minimum width=3.4cm, minimum height=1.1cm, align=center},
      outputbox/.style={rectangle, rounded corners, draw=black!70!black, fill=white!10, thick, minimum width=3cm, minimum height=1cm, align=center},
      arrow/.style={-{Latex[length=2.5mm]}, thick, black!80},
      grouplabel/.style={font=\small\bfseries, text=black!70},]
    
    \node[inputbox] (input) {
      \textbf{TNG50-1 / TNG100-1 cutout} \\[1pt]
      \footnotesize HDF5 stellar particles \\[-1pt]
      \footnotesize ($x, y, z, v, m$)
    };

    \node[processbox, below=of input] (center) {
      \textbf{1. Shrinking-sphere centre} \\[1pt]
      \footnotesize periodic unwrap;~\citep{2003MNRAS.338...14P} \\[-1pt]
      \footnotesize $R_c \to 0.8 R_c$ until $R_c<0.5\,\mathrm{kpc}$ or $N<128$
    };

    \node[processbox, below=of center] (velocity) {
      \textbf{2. Velocity conversion} \\[1pt]
      \footnotesize $v_{\rm phys} = \sqrt{a}\; v_{\rm code}$ \\[-1pt]
      \footnotesize (Springel 2005; Nelson+ 2019)
    };

    \node[processbox, below=of velocity] (orient) {
      \textbf{3. Face-on orientation} \\[1pt]
      \footnotesize Diagonalise $\mathbf{L}$ inside $10\,\mathrm{kpc}$; \\[-1pt]
      \footnotesize rotate so $\hat{L}\parallel\hat z$
    };

    \node[processbox, below=of orient] (cut) {
      \textbf{4. Disc particle selection} \\[1pt]
      \footnotesize Keep $|z|<3\,\mathrm{kpc}$, $R<30\,\mathrm{kpc}$ \\[-1pt]
      \footnotesize (skip if $N_{\rm disc}<500$)
    };

    \node[processbox, right=3.1cm of orient, yshift=6.2cm] (fourier) {
      \textbf{5. Fourier decomposition} \\[1pt]
      \footnotesize $c_m(R)=\sum_i m_i e^{im\phi_i}$ \\[-1pt]
      \footnotesize $A_m(R)=2|c_m|/c_0$
    };

    \node[processbox, below=of fourier] (global) {
      \textbf{6. Global amplitude} \\[1pt]
      \footnotesize $\langle A_m\rangle$ over $[1.5,\,2.5]\,R_d$ \\[-1pt]
      \footnotesize (R\&Z95 + Jog integral)
    };

    \node[decisionbox, below=of global] (bar) {
      \textbf{7. Bar length \& coherence} \\[1pt]
      \footnotesize Phase-coherent extent~\citep{2009A&A...495..491A}; \\[-1pt]
      \footnotesize flag $f_{\rm coh}>0.5$
    };

    \node[decisionbox, below=of bar] (boot) {
      \textbf{8. Uncertainties \& noise floor} \\[1pt]
      \footnotesize bootstrap, $N_{\rm boot}=200$~\citep{Efron1995AnIT} \\[-1pt]
      \footnotesize azimuth shuffle $\to A_{1,\rm noise}$, $A_{1,\rm corr}$
    };

    \node[outputbox, below=of boot] (output) {
      \textbf{Catalog row} \\[1pt]
      \footnotesize \texttt{catalog\_v2.csv} \\[-1pt]
      \footnotesize (45 columns per galaxy)
    };

    \draw[arrow] (input) -- (center);
    \draw[arrow] (center) -- (velocity);
    \draw[arrow] (velocity) -- (orient);
    \draw[arrow] (orient) -- (cut);
    \draw[arrow]
  (cut.east) -- ++(1.4,0)
  |- ([yshift=0.07cm]fourier.west);
    \draw[arrow] (fourier) -- (global);
    \draw[arrow] (global) -- (bar);
    \draw[arrow] (bar) -- (boot);
    \draw[arrow] (boot) -- (output);

    \begin{scope}[on background layer]
      \node[fit={(center) (cut)}, rounded corners, draw=black!30, dashed, thick, inner sep=6pt] (pre) {};
      \node[grouplabel, anchor=east] at ([xshift=-8pt] pre.west)
            {Preprocessing};

      \node[fit={(fourier) (bar)}, rounded corners, draw=black!30, dashed, thick, inner sep=6pt] (analysis) {};
      \node[grouplabel, anchor=west] at ([xshift=8pt] analysis.east)
            {Fourier analysis};

      \node[fit={(boot) (output)}, rounded corners, draw=black!30, dashed, thick, inner sep=6pt] (out) {};
      \node[grouplabel, anchor=west] at ([xshift=8pt] out.east)
            {Uncertainty \& output};
    \end{scope}

  \end{tikzpicture}

  \caption{Overview of the \pipeline\ processing pipeline.  A raw HDF5 particle cutout from TNG50-1 or TNG100-1 is first unwrapped across the periodic box and enters the preprocessing block (Steps 1--4), which delivers a centred, face-on, disc-restricted particle set to the Fourier analysis block (Steps 5--7).  Step 8 estimates bootstrap uncertainties on all reported quantities and the shot-noise floor $A_{1,\rm noise}$ from azimuth-randomised copies of the disc, from which the noise-corrected amplitude $A_{1,\rm corr}$ is obtained (Section \ref{sec:noise_floor}); the results are appended as a single row of the catalogue.  Each stage is described in detail in the corresponding subsection of Section \ref{sec:algorithm}.}
  \label{fig:pipeline_flowchart}
\end{figure*}
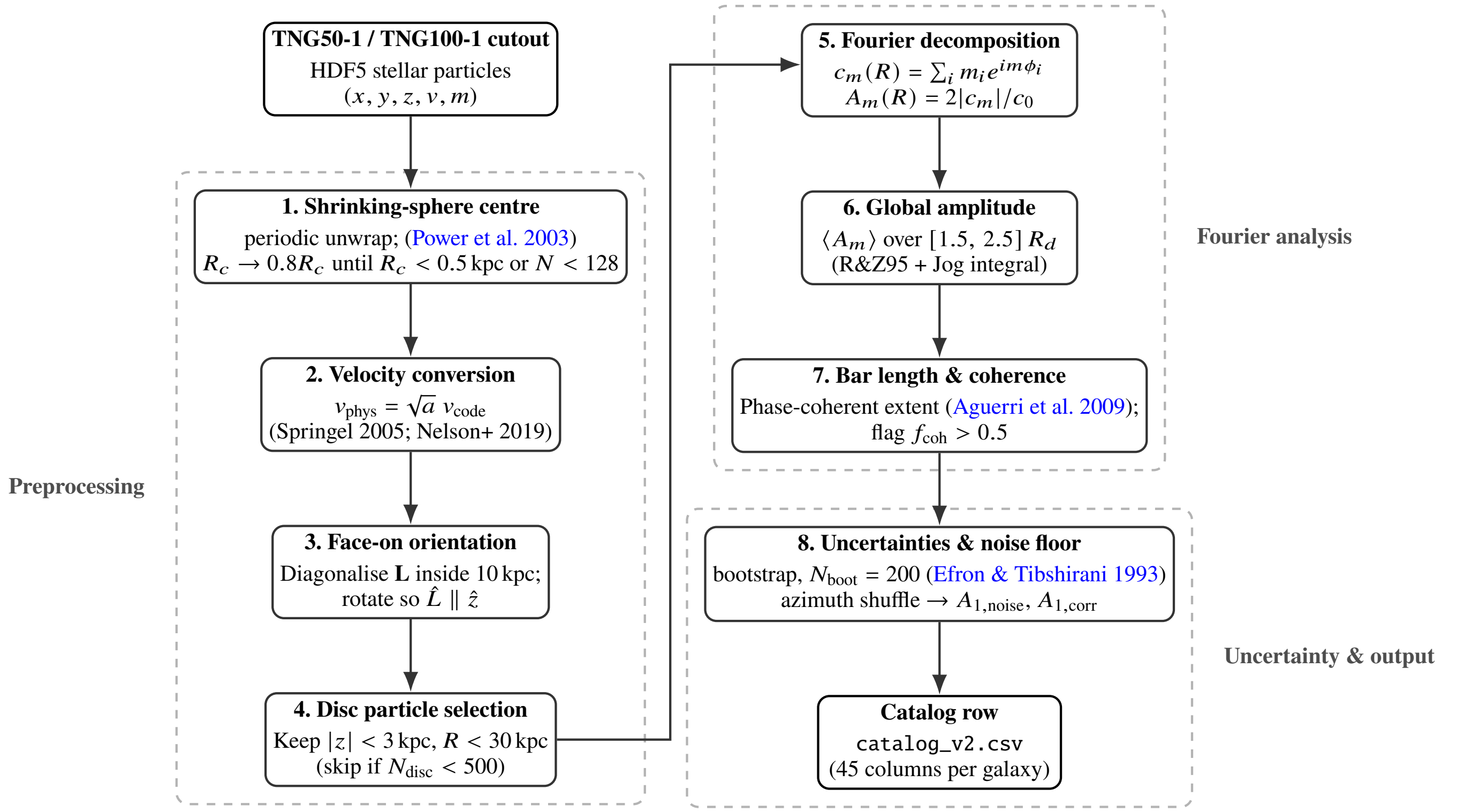

\subsection{Fourier decomposition normalisation}
\label{sec:fourier_norm}

Figure \ref{fig:pipeline_flowchart} summarises the eight-stage processing pipeline described in this section, from raw HDF5
cutout to catalogue row. Following R\&Z95, we compute the complex Fourier coefficients $c_m(R)$ in radial bins of width $\Delta R$ from the particle distribution:
\begin{equation}
  c_m(R) = \sum_{i \in \Delta R} w_i \, m_i \, e^{\mathrm{i} m \phi_i},
  \label{eq:cm_particle}
\end{equation}
where $m_i$ is the particle mass and $w_i$ is an optional weight (default $w_i = 1$).  With this sign convention the phase $\phi_m$ in Eq. \ref{eq:fourier} is the azimuth of the density maximum of mode $m$.  The amplitude and phase are
\begin{equation}
  \Am{m}(R) = \frac{2\,|c_m(R)|}{c_0(R)}, \qquad
  \phi_m(R) = \frac{1}{m}\arg c_m(R),
  \label{eq:Am_phim}
\end{equation}
where the factor of 2 ensures that $\Am{m}$ equals the fractional amplitude of the corresponding sinusoidal perturbation in $\Sigma(R,\phi)$; see Appendix \ref{app:norm}.  This is the normalisation adopted by R\&Z95, \citet{2005A&A...438..507B},
\citet{2013ApJ...772..135Z}, and most subsequent lopsidedness catalogues.

\subsection{Global amplitude in an aperture}

Following R\&Z95 (their Eq. 5) we report an aperture-averaged amplitude,
\begin{equation}
  \langle \Am{m} \rangle_{\rm lit} = \frac{1}{R_{\rm out} - R_{\rm in}}
      \int_{R_{\rm in}}^{R_{\rm out}} \Am{m}(R)\,dR,
  \label{eq:literature_average}
\end{equation}
with default aperture $[1.5, 2.5]\,\Rd$, following the standard established by \citet{1995ApJ...447...82R} and adopted by \citet{1997ApJ...477..118Z} and \citet{2013ApJ...772..135Z}. This aperture range excludes the bulge-dominated inner disc (where non-exponential profiles complicate the aperture average) and the shot-noise-dominated outer disc, providing the optimal signal-to-noise regime for the global $m=1$ measurement.  Alternatively, the fully mass-weighted \citet{2002A&A...391..471J} form,
\begin{equation}
  \langle \Am{m} \rangle_{\Sigma} =
     \frac{\int \Am{m}(R)\,\Sigma_0(R)\,2\pi R\,dR}
          {\int \Sigma_0(R)\,2\pi R\,dR},
  \label{eq:jog_integral}
\end{equation}
is optionally supported when $\Sigma_0(R)$ is supplied.

\begin{algorithm}[t]
  \caption{Particle-based Fourier decomposition (\textsc{compute\_fourier})}
  \label{alg:fourier}
  \DontPrintSemicolon
  \KwIn{Particle positions $(x_i, y_i)$, masses $m_i$; radial range $[R_{\rm in}, R_{\rm out}]$; number of bins $N_{\rm bin}$; maximum mode $m_{\rm max}$}
  \KwOut{Radial profiles $A_m(R)$, $\phi_m(R)$ for $m = 1, \dots, m_{\rm max}$}
  \BlankLine
  Convert to polar: $R_i \leftarrow \sqrt{x_i^2 + y_i^2}$, $\phi_i \leftarrow \arctan2(y_i, x_i)$\;
  Define radial bin edges $R_k$ for $k = 0, \dots, N_{\rm bin}$ with uniform spacing in $[R_{\rm in}, R_{\rm out}]$\;
  \ForEach{bin $k$}{ Identify particles with $R_k \le R_i < R_{k+1}$\;
    Compute $c_0(R_k) \leftarrow \sum_{i \in k} m_i$\;
    \If{$c_0(R_k) = 0$}{
      Mark bin as empty; continue\;
    }
    \ForEach{$m = 1, \dots, m_{\rm max}$}{
      $c_m(R_k) \leftarrow \sum_{i \in k} m_i \exp(\mathrm{i} m \phi_i)$\;
      $A_m(R_k) \leftarrow 2\,|c_m(R_k)| / c_0(R_k)$\;
      $\phi_m(R_k) \leftarrow \arg[c_m(R_k)] / m$\;
    }
  }
  \Return $\{A_m(R_k), \phi_m(R_k)\}$
\end{algorithm}

\subsection{Bar length: phase-coherent definition}
\label{sec:bar_length}

Naive amplitude-threshold estimators ($L_{\rm bar} = \max\{R : \Am{2}(R) > 0.2\}$) suffer from shot-noise contamination in the outer disc.  Bar identification via the radial extent of the $m=2$ Fourier amplitude follows a long-established tradition \citep{1990ApJ...357...71O,2002MNRAS.330...35A,2005MNRAS.364..283E}. Following \citet{2005A&A...434..109A} and \citet{2009A&A...495..491A}, we adopt a phase-coherence criterion.  The resulting $L_{\rm bar}$ is a heuristic: it is the radial extent of a phase-coherent $m=2$ pattern on the adopted grid, which may include a spiral extension in phase with the bar, and it is not a calibrated bar length.  Let $R_{\rm peak}$ be the radius of the $\Am{2}$ maximum, with phase $\phi_2^{\rm ref} = \phi_2(R_{\rm peak})$.  The extent $L_{\rm bar}$ is the largest radius $R$ satisfying
\begin{equation}
  \Am{2}(R') > \Am{2, \min} \quad \text{and} \quad
  |\phi_2(R') - \phi_2^{\rm ref}|_{\pi} < \Delta\phi_{\rm tol}
  \label{eq:bar_length}
\end{equation}
for all $R_{\rm peak} \le R' \le R$, with defaults $\Am{2,\min} = 0.2$ and $\Delta\phi_{\rm tol} = 22.5^{\circ}$.  In parallel we report a \emph{pattern coherence flag},
\begin{equation}
  f_{\rm coh} = \frac{
    \#\{R : \Am{2}(R) > \Am{2,\min}\,\text{and}\,
    |\phi_2(R) - \bar\phi_2|_{\pi} < \Delta\phi_{\rm tol}\}}{
    \#\{R : \Am{2}(R) > \Am{2,\min}\}},
  \label{eq:fcoh}
\end{equation}
where $\bar\phi_2 = \tfrac{1}{2}\arg\sum_R A_2(R)\,e^{2{\rm i}\phi_2(R)}$ is the amplitude-weighted mean phase and $|x|_\pi$ denotes a phase difference reduced to $(-\pi/2, \pi/2]$, since $\phi_2$ has period $\pi$.  Detections with $f_{\rm coh} < 0.5$ are treated as non-detections.  For our TNG50 sample, 66.7\% of galaxies (610/915) exhibit $f_{\rm coh} > 0.5$; the sensitivity of this fraction to the threshold is shown in Section \ref{sec:parameter_sensitivity}.

When the phase-coherent $m=2$ structure extends to the outermost analysed radial bin, the reported bar length corresponds to the
analysis boundary and should therefore be interpreted as a lower limit rather than a measured termination radius.  The radial
Fourier profiles are computed on a linear grid of $n_{\rm bins} = 40$ bins spanning $[0, r_{\rm out}]$ with $r_{\rm out} = 15\kpc$ (bin width $0.375\kpc$), so the outermost bin centre -- and hence the maximum recoverable $L_{\rm bar}$ -- is at $14.8125\kpc$. In the TNG50 catalogue this saturation occurs in $36.0\%$ of galaxies (329/915), where the recovered $L_{\rm bar}$ equals the outermost bin centre; these values are flagged in the machine-readable catalogue as boundary-limited.  The saturated fraction is dominated by discs in which a strong bar and its outer spiral extension share a common phase, so that the coherence criterion is satisfied throughout the adopted radial grid.  Statistical use of $L_{\rm bar}$ in this paper is restricted to the detection dichotomy $f_{\rm coh} > 0.5$ vs.\ $f_{\rm coh} \le 0.5$; no distribution of bar-length values is
reported for astrophysical inference.

\begin{algorithm}[t]
  \caption{Phase-coherent bar length (\textsc{bar\_length})}
  \label{alg:bar_length}
  \DontPrintSemicolon
  \KwIn{Radial profile $A_2(R)$, $\phi_2(R)$; thresholds $A_{2,\min}$, $\Delta\phi_{\rm tol}$}
  \KwOut{Bar length $L_{\rm bar}$; coherence flag $f_{\rm coh}$}
  \BlankLine
  Locate peak: $R_{\rm peak} \leftarrow \arg\max_R A_2(R)$ subject to $A_2(R) > A_{2,\min}$\;
  \If{no $R$ satisfies $A_2 > A_{2,\min}$}{
    \Return $L_{\rm bar} \leftarrow 0$, $f_{\rm coh} \leftarrow 0$\;
  }
  Set reference phase $\phi_2^{\rm ref} \leftarrow \phi_2(R_{\rm peak})$\;
  Initialise $L_{\rm bar} \leftarrow R_{\rm peak}$\;
  \tcp{Extend outward while amplitude and phase are coherent}
  \ForEach{$R > R_{\rm peak}$ in increasing order}{
    \eIf{$A_2(R) > A_{2,\min}$
         \KwSty{ and }
         $|\phi_2(R) - \phi_2^{\rm ref}|_{\pi} < \Delta\phi_{\rm tol}$}{
      $L_{\rm bar} \leftarrow R$\;
    }{
      break\;
    }
  }
  \BlankLine
  \tcp{Coherence quality flag over all radii with $A_2 > A_{2,\min}$}
  $\bar\phi_2 \leftarrow \tfrac{1}{2}\arg\sum_R A_2(R)\,e^{2\mathrm{i}\phi_2(R)}$\;
  $N_{\rm active}  \leftarrow |\{R : A_2(R) > A_{2,\min}\}|$\;
  $N_{\rm coherent} \leftarrow |\{R : A_2(R) > A_{2,\min}
                              \KwSty{ and } |\phi_2(R) - \bar\phi_2|_{\pi} < \Delta\phi_{\rm tol}\}|$\;
  $f_{\rm coh} \leftarrow N_{\rm coherent} / N_{\rm active}$\;
  \Return $L_{\rm bar}, f_{\rm coh}$
\end{algorithm}

\subsection{Iterative centring and disc orientation}

For cosmological snapshots we implement the shrinking-sphere algorithm of \citet{2003MNRAS.338...14P}, which proceeds as follows:
\begin{enumerate}
  \item Initialise the centre (seed) at the position of the minimum-potential star particle.
  \item Compute the centre of mass within a sphere
        of radius $R_c$ (default $R_c = 30\kpc$).
  \item Shrink $R_c \to 0.8\,R_c$ and iterate until either
        $R_c < 0.5\kpc$ or $<128$ particles remain.
\end{enumerate}
The disc plane is defined by the angular-momentum vector of star particles within $10\kpc$ of the centre; the coordinate frame is rotated so that this vector aligns with $\hat{z}$.  Only particles with $|z| < 3\kpc$ and $R < 30\kpc$ are used in the subsequent Fourier decomposition.

Before centring, particle coordinates are unwrapped across the periodic boundary of the simulation box about the minimum-potential particle.  Without this step a galaxy close to the box edge loses particles to the opposite side of the box.  In our samples 17 TNG50-1 and 20 TNG100-1 galaxies (1.9 and 0.3 per cent) have particle sets that straddle the boundary; in all of them the displaced particles belong to the outer stellar halo, and re-processing them after unwrapping changes $\Amone$ by less than
$10^{-4}$ and $D/T$ by less than 0.5 per cent.

\subsection{Velocity unit correction for IllustrisTNG}
\label{sec:sqrt_a}

IllustrisTNG snapshots store peculiar velocities in the internal GADGET-2 code convention \citep[][and IllustrisTNG data-release documentation]{2005MNRAS.364.1105S,2019ComAC...6....2N}: $\mathbf{v}_{\rm peculiar} = \sqrt{a}\,\mathbf{v}_{\rm code}$, where $a$ is the cosmic scale factor of the snapshot.  Physical peculiar velocities in $\kms$ are recovered by multiplying by $\sqrt{a}$ (a no-op at $z=0$ but essential at higher redshift). Failing to apply this correction biases angular-momentum estimates and misorients disc frames.

\subsection{Bootstrap uncertainties}
\label{sec:bootstrap}

We estimate statistical uncertainties by resampling the disc particles with replacement \citep{Efron1995AnIT}, with $\Nboot = 200$ iterations (Algorithm \ref{alg:bootstrap}).  Typical standard errors on aperture-averaged amplitudes are $\sigma_{\Am{m}} \sim 0.005$ for $\Ndisk \sim 10^5$.  The bootstrap standard deviation is a good estimate of the statistical error. The bootstrap \emph{mean}, however, is not an unbiased estimate of the amplitude: each resample adds its own Poisson noise to a quantity that is already biased upwards (Section \ref{sec:noise_floor}), so $\langle \hat A_m\rangle_{\rm boot} > A_m$ on average.  The
catalogue reports the bootstrap mean for continuity with earlier versions.  The quantitative TNG50-1/TNG100-1 comparisons of Sections \ref{sec:comparison_lokas} and \ref{sec:resolution} are based on the direct estimate of Eq. \ref{eq:literature_average} ($A_1^{\rm lit}$) and the noise-corrected amplitude defined below; the bootstrap mean is shown alongside them for reference, and is still used in the $\Amone < 0.5$ sample cut, in Figs \ref{fig:fig9_rotation_analysis} and \ref{fig:fig6_obs_comparison}, and in the catalogue excerpt of Appendix \ref{app:catalog_columns}. We also list the standard bias-corrected bootstrap estimate
$\hat A_m^{\rm bc} = 2A_m^{\rm lit} - \langle \hat A_m\rangle_{\rm boot}$ \citep{Efron1995AnIT} as an independent check.

\begin{algorithm}[t]
  \caption{Bootstrap uncertainty on global amplitude
    (\textsc{bootstrap\_fourier})}
  \label{alg:bootstrap}
  \DontPrintSemicolon
  \KwIn{Disc particles $\{(x_i, y_i, m_i)\}$ with $N$ entries; number of iterations $N_{\rm boot}$; target mode $m$}
  \KwOut{Bootstrap mean $\hat A_m$, standard error $\sigma_{\hat A_m}$}
  \BlankLine
  Initialise empty list $\mathcal{L} \leftarrow \emptyset$\;
  \For{$b = 1, \dots, N_{\rm boot}$}{
    Draw $N$ indices $\{j_1, \dots, j_N\}$ uniformly with replacement from $\{1, \dots, N\}$\;
    Form the resampled particle set
       $\{(x_{j_i}, y_{j_i}, m_{j_i})\}_{i=1}^N$\;
    Run Algorithm \ref{alg:fourier} on the resampled set\;
    Compute the aperture-averaged amplitude
       $\langle A_m\rangle_b$ over $[1.5, 2.5]\,R_{\rm d}$
       following Eq. (\ref{eq:literature_average})\;
    Append $\langle A_m\rangle_b$ to $\mathcal{L}$\;
  }
  $\hat A_m \leftarrow \operatorname{mean}(\mathcal{L})$\;
  $\sigma_{\hat A_m} \leftarrow \operatorname{std}(\mathcal{L})$\;
  \tcp{Bias-corrected estimate (Section \ref{sec:bootstrap})}
  $\hat A_m^{\rm bc} \leftarrow 2A_m^{\rm lit} - \hat A_m$\;
  \Return $\hat A_m, \sigma_{\hat A_m}$\;
\end{algorithm}

\subsection{Shot-noise floor and noise-corrected amplitudes}
\label{sec:noise_floor}

For $N$ equal-mass particles placed at random azimuths in an annulus, $|c_m|/c_0$ follows a Rayleigh distribution and the expected amplitude in the R\&Z95 normalisation is $\langle A_m\rangle_{\rm noise} \simeq \sqrt{\pi/N}$.  For unequal masses $N$ is replaced by the effective number $N_{\rm eff} = (\sum m_i)^2/\sum m_i^2$.  A measured amplitude therefore contains a shot-noise contribution
whose expected magnitude is $\langle A_m\rangle_{\rm noise}$, so two galaxies with the same true $A_1$ but different particle numbers have different expected measurements.

We estimate the floor empirically for each galaxy.  The azimuths of the disc particles are replaced by random values while their radii and masses are kept, which removes all non-axisymmetric structure but preserves the radial distribution of particles.  The resulting profile is computed five times on the same radial grid and averaged, and its aperture mean defines $A_{1,\rm noise}$.  The noise-corrected amplitude is
\begin{equation}
  A_{1,\rm corr} = \sqrt{\max\!\left[(A_1^{\rm lit})^2 -A_{1,\rm noise}^2,\; 0\right]},
  \label{eq:a1corr}
\end{equation}
the usual quadrature subtraction of a noise term from a positive-definite amplitude.  The same correction is applied in two fixed physical apertures, 3.5--6 and 5--8\,kpc, which are also listed in the catalogue.

Equation (\ref{eq:a1corr}) is an approximation, and its behaviour can be made explicit. In a single annulus the noise term of $2c_1/c_0$ is approximately a complex Gaussian variable with variance $\sigma^2$ per mode, so for a true amplitude $s$ the measured amplitude follows a Rice distribution, with $\langle A^2\rangle = s^2 + \langle A_{\rm noise}^2\rangle$ exactly, whereas $\langle A_{\rm noise}\rangle^2 = (\pi/4)\langle A_{\rm noise}^2\rangle$.  Subtracting the squared \emph{mean} amplitudes, as in Eq. (\ref{eq:a1corr}), is therefore exact for $s = 0$, where $\langle A\rangle = \langle A_{\rm noise}\rangle$, but not in general: for $s \gg \sigma$ one has $\langle A\rangle \simeq s + \sigma^2/(4s)$, so that $\langle A\rangle^2 - \langle A_{\rm noise}\rangle^2 \simeq s^2 - (\pi/4 - 1/2)\sigma^2 \approx s^2 - 0.36\,\langle A_{\rm noise}\rangle^2$, with $\sigma^2 = \langle|n|^2\rangle$ the noise variance of the complex coefficient.  The correction thus slightly over-subtracts at high signal-to-noise, by a fractional amount $\approx 0.18\,(A_{1,\rm noise}/A_1)^2$ in the amplitude:
$\approx 2$ per cent for a median TNG50-1 disc and $\approx 5$--6 per cent for a median TNG100-1 disc.  This bias lowers the corrected TNG100-1 amplitudes more than the TNG50-1 ones, so it can only make the TNG100/TNG50 excess of Section \ref{sec:resolution} appear smaller, not larger.

The power-subtraction alternative, $\sqrt{\langle A^2\rangle - \langle A_{\rm noise}^2\rangle}$ evaluated over the aperture bins, is unbiased in $s^2$ bin by bin.  We compared the two estimators on synthetic lopsided discs ($R_{\rm d} = 3\kpc$, $\varepsilon_1 = 0$, 0.03, 0.05, 0.10 and 0.20, $N = 5\times10^3$--$5\times10^5$, eight realisations each; script \texttt{noise\_estimator\_test.py}).  The mean recovered amplitudes of the two estimators differ by less than 5 per cent of $\varepsilon_1$ for $\varepsilon_1 \ge 0.05$ at every particle number, and their scatter is the same.  Both are reliable when there are $\gtrsim 700$ particles per aperture bin: for $\varepsilon_1 = 0.10$ they recover 0.105--0.110 at $\sim 700$ particles per bin and 0.100 at $\sim 3400$.  Below $\sim 200$ particles per bin both become unreliable: an axisymmetric disc gives a residual of $\simeq 0.03$--0.04, and an injected $\varepsilon_1 = 0.10$ is recovered as $0.07 \pm 0.05$.  The filtered TNG100-1 discs have 400--950 effective particles per aperture bin (Table \ref{tab:downsampling}), between the two regimes of the synthetic test; the downsampling test shows that even at $N_{\rm eff}\approx 290$ the median $A_{1,\rm corr}$ is recovered to
3 \% (Table \ref{tab:downsampling}).  The lowest-mass downsampled TNG50-1 bin (150 per bin) lies in the unreliable regime,
and we do not use it in the comparison.  We also test the correction directly in Section \ref{sec:resolution} by downsampling
TNG50-1 discs to the particle mass of TNG100-1.  For the median TNG50-1 disc the correction is small ($A_{1,\rm noise}/A_1^{\rm lit} \approx 0.3$, with $N_{\rm eff}\approx 3\times10^4$ in the aperture); for TNG100-1 it is not ($A_{1,\rm noise}/A_1^{\rm lit} \approx 0.5$--0.6, with $N_{\rm eff}\approx 2$--$4\times10^3$).

\subsection{Runtime and reproducibility}
\label{sec:runtime}
\begin{figure*}
  \centering
  \includegraphics[width=\textwidth]{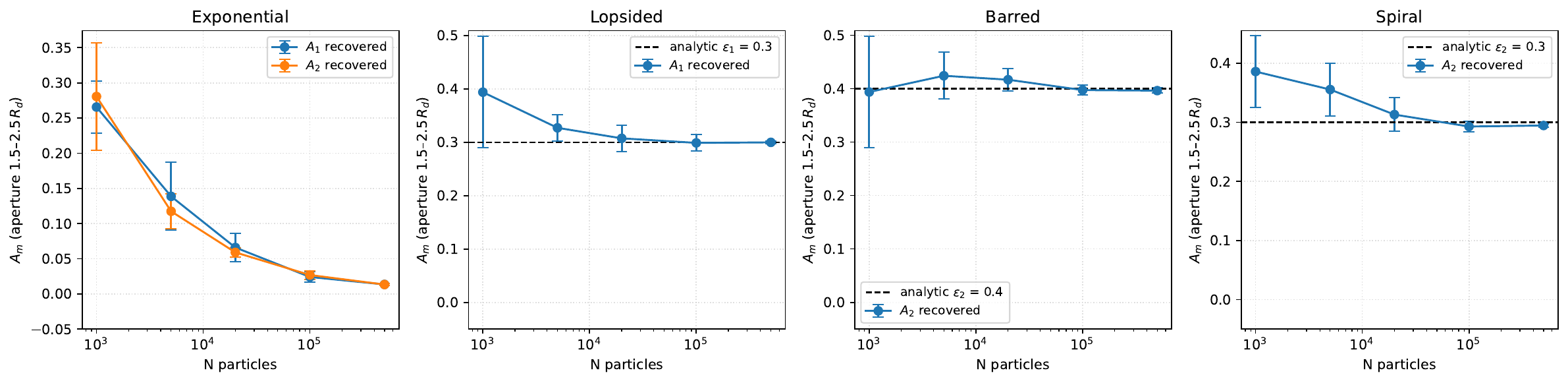}
  \caption{Recovery of injected Fourier amplitudes as a function of particle count for four synthetic galaxy families ($\Rd = 3\kpc$, aperture $[1.5, 2.5]\,\Rd$, five random realisations per point).  Dashed lines: analytic amplitudes. All three perturbed galaxies converge to within $<2\%$ at $N = 5\times10^5$.}
  \label{fig:fig1_amplitude_recovery}
\end{figure*}
The pipeline is designed for straightforward reproducibility on commodity hardware.  All results in this paper were produced on a
single Windows workstation with Anaconda Python 3.12.7 (no GPU acceleration; single-threaded execution).  Table \ref{tab:runtime}
lists wall-clock timings for the four batch pipeline stages applied to the complete parent sample of 923 TNG50-1 subhalos. All 923 are processed; 915 of them pass the requirement of at least 500 disc particles (Section   \ref{sec:sample_selection}) and form the sample analysed in the rest of the paper.

\begin{table}
  \centering
  \caption{Runtime and data volume for the complete 923-galaxy TNG50-1 batch analysis, measured on a single Windows/Anaconda workstation.  The pipeline is resumable: failed or interrupted stages skip already-completed subhalos.  The batch contains all 923 candidates, of which 915 pass the disc-particle requirement.  The total wall time exceeds the sum of stage times by $\sim 1.2\,\mathrm{h}$ because of I/O overhead (HDF5 open/close cycles, disk seeks), TNG API retry delays for a small fraction of subhalos, and inter-stage disk flushes not included in the individual stage timings.}
  \label{tab:runtime}
  \begin{tabular}{lrl}
    \toprule
    Pipeline stage             & Time    & Data volume \\
    \midrule
    (1) Query TNG API          & $\sim$1 min   & --- \\
    (2) Download HDF5 cutouts  & $\sim$3\,h    & 13.1\,GB \\
    (3) Fourier processing     & $\sim$1.5\,h  & 1.1\,MB (JSON) \\
    (4) Build CSV / FITS catalogue & $<30$\,s  & 210\,kB (CSV) \\
    \midrule
    Total wall time            & \textbf{5.76\,h}  &  \\
    Per-galaxy processing      & $\sim$22.5\,s  & \\
    \bottomrule
  \end{tabular}
\end{table}

The dominant cost is the download step, which is I/O-bound and rate-limited to be considerate to the TNG API servers ($\sim$1.5\,s between requests); the actual computation is trivially parallelisable by launching multiple instances of stage 3 (Fourier processing) on non-overlapping subhalo subsets.  On a modern multi-core node with local access to the raw snapshot data, we estimate that the entire pipeline can be reduced to under one hour for the 923-galaxy sample.

To facilitate independent reproduction of our results, the \pipeline\ code repository includes a pinned dependency specification (\texttt{requirements.txt} for pip and \texttt{environment.yml} for conda), a \texttt{Dockerfile} for container-based execution on any system, and a continuous integration workflow that automatically re-runs the synthetic-galaxy validation on every commit (GitHub Actions, tested on Linux/macOS/Windows with Python 3.11 and 3.12).  The full analysis of Section \ref{sec:tng50} can be
reproduced by cloning the repository, editing \texttt{batch\_tng50/batch\_config.yaml} to add a valid TNG API token, and running the four batch pipeline stages in sequence.

\section{Validation on synthetic galaxies}
\label{sec:validation}

\subsection{Synthetic galaxy families}

We construct four synthetic galaxy families with analytically prescribed Fourier amplitudes: an axisymmetric \textbf{Exponential}
disc (expected $\Am{m} = 0$), a \textbf{Lopsided} disc with $\varepsilon_1 = 0.3$, a \textbf{Barred} disc with $\varepsilon_2 = 0.4$, and a \textbf{Logarithmic Spiral} with pitch angle 15$^\circ$ and effective $\varepsilon_s = 0.3$. All have $\Rd = 3\kpc$, sampled at $N = 10^3$ to $5\times10^5$ particles with five random realisations per point.

\subsection{Recovery of injected amplitudes}
\label{sec:validation_recovery}

Figure \ref{fig:fig1_amplitude_recovery} shows the recovered $\Am{m}$ values as a function of particle count $N$, for the literature aperture $[1.5, 2.5]\,\Rd$.  At $N = 5\times10^5$:
\begin{itemize}
  \item Lopsided: $\Amone = 0.2998 \pm 0.0020$
        ($-0.07\%$ bias vs.\ $\varepsilon_1 = 0.30$);
  \item Barred: $\Amtwo = 0.3963 \pm 0.0049$
        ($-0.93\%$ bias vs.\ $\varepsilon_2 = 0.40$);
  \item Spiral: $\Amtwo = 0.2946 \pm 0.0032$
        ($-1.8\%$ bias vs.\ $\varepsilon_s = 0.30$).
\end{itemize}
The axisymmetric exponential galaxy shows a residual $\Am{m} \sim 0.014$ consistent with shot noise.

\subsection{Shot-noise scaling}
\label{sec:shot_noise}

Figure \ref{fig:fig2_shot_noise_scaling} (Appendix \ref{app:supplementary}) tests whether the across-realisation scatter follows the expected shot-noise scaling $\sigma(\Am{m}) \propto N^{-1/2}$.  All four galaxy families follow the expected $N^{-1/2}$ slope, lying at or below the conservative reference scale $\sigma_{\rm theory} = \sqrt{2\pi / N_{\rm aperture}}$ across three
orders of magnitude in $N$.  The shaded band represents the 95\% confidence interval expected for five realisations under a $\chi^2$ distribution, and encloses all data points.  The radial profiles at $N = 10^5$ recover the analytic amplitudes across the disc, above a shot-noise floor of $\sim$0.015 at $N_{\rm ap} \sim 2.8\times10^4$.

\subsection{Extended validation: inclination, centring, and particle loss}
\label{sec:extended_validation}

The tests in Sections \ref{sec:validation_recovery}--\ref{sec:shot_noise} demonstrate that the pipeline correctly recovers analytic Fourier amplitudes in a face-on, fully-sampled, perfectly-centred synthetic galaxy.  Real cosmological analyses, however, must contend with three additional systematic effects: (i) non-zero disc inclination that biases the projected surface density; (ii) small offsets between the adopted centre and the true dynamical centre; and (iii) stochastic loss of particles from mass cuts, star-formation history selection, or numerical artefacts.  We characterise each of these effects using a reference lopsided disc with $\varepsilon_1 = 0.30$ and $\Rd = 3\kpc$ at $N = 10^5$ particles, running 5 random realisations at each systematic level.

\begin{figure*}
  \centering
  \includegraphics[width=\textwidth]{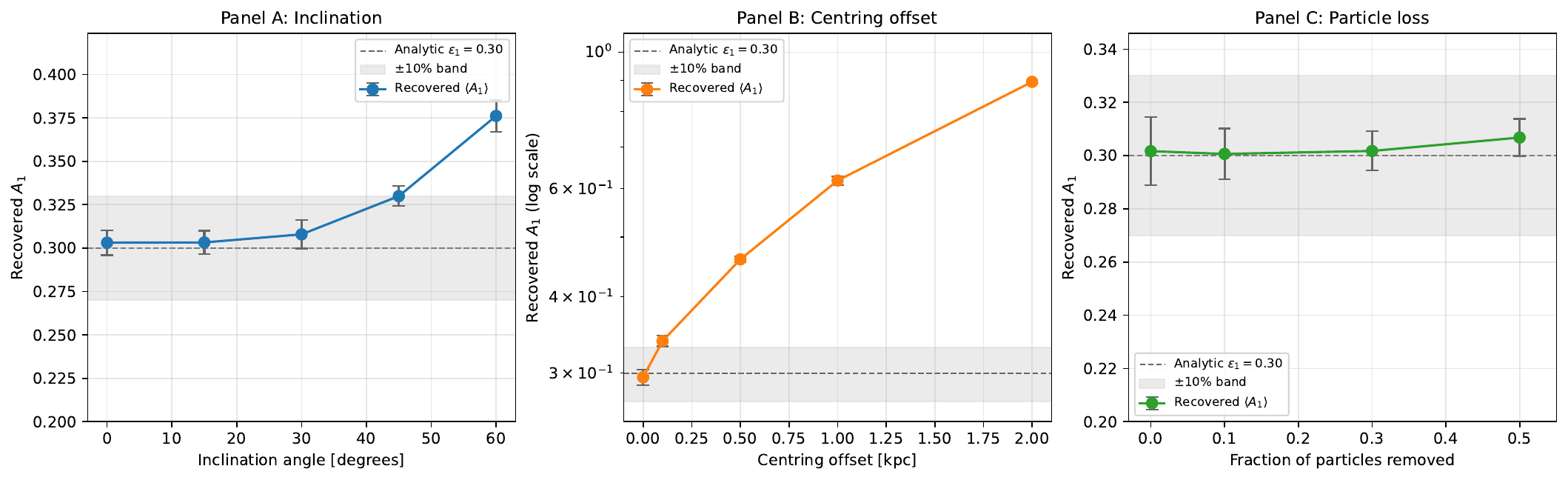}
  \caption{Extended systematic-effect validation of the \pipeline\ pipeline on a reference lopsided synthetic disc ($\varepsilon_1 = 0.30$, $\Rd = 3\kpc$, $N = 10^5$ particles, 5 realisations per level).  \textbf{Panel A}: recovered $\Amone$ vs.\ disc inclination angle.  \textbf{Panel B}: recovered $\Amone$ vs.\ a forced offset applied to the pipeline centre. \textbf{Panel C}: recovered $\Amone$ after removing a random fraction of the particles.  Dashed grey line: injected $\varepsilon_1 = 0.30$; grey band: $\pm 10\%$ tolerance.  Error bars are 1$\sigma$ across five realisations.}
  \label{fig:fig7_extended_validation}
\end{figure*}

Figure \ref{fig:fig7_extended_validation} summarises the outcome. Panel A shows that the recovered $\Amone$ is robust against
inclination for tilts $i \lesssim 30^{\circ}$, with $|\Delta \Amone / \Amone| < 3\%$; at $i = 45^{\circ}$ the bias reaches $\sim 10\%$, and at $i = 60^{\circ}$ it grows to $\sim 25\%$.  This behaviour is consistent with the geometric expectation that inclination-induced projection distortions become important once $\cos i \lesssim 0.7$, motivating a face-on selection cut ($i \lesssim 30^{\circ}$) for precision lopsidedness studies.

Panel B reveals the \emph{dominant} systematic in our validation suite: centring accuracy.  Shifting the assumed centre by even $0.1\kpc$ produces a $\sim 13\%$ spurious enhancement of $\Amone$; a $0.5\kpc$ offset raises the recovered amplitude by $50\%$, and a $1\kpc$ offset \emph{doubles} it (recovered $\Amone \simeq 0.62$ vs.\ injected $0.30$).  This steep sensitivity arises because an off-centre mass distribution mimics an $m=1$ perturbation by construction: the pipeline's Fourier decomposition cannot distinguish a genuine lopsided disc from a symmetric disc displaced from the analysis origin.  The result underscores the importance of robust centring algorithms for any Fourier-based lopsidedness study. For our TNG50 analysis, we use the iterative shrinking-sphere algorithm of \citet{2003MNRAS.338...14P}.  Since a 0.1\,kpc offset already raises $\Amone$ by $\sim 13$ per cent in this test, centring can contribute a systematic uncertainty of order 10 per cent and is potentially the dominant pipeline systematic.  The forced-offset test does not measure how accurate the centring actually is in simulated galaxies.  An indirect check is provided by the downsampling test of
Section \ref{sec:resolution}: when 94 per cent of the star particles of a filtered TNG50-1 disc are removed and the centre is recomputed, it moves by a median of 0.03\,kpc (84th percentile 0.07\,kpc, 95th percentile 0.14\,kpc), and the corrected amplitudes agree with the full-resolution ones to within a few per cent in the median.  This shows that the centre is stable, not that it coincides with the true dynamical centre; a comparison with an independent centre definition (e.g.\ the dark-matter potential minimum) is left for future work.
Panel C tests robustness to random particle loss.  Removing up to 50\% of the particles at random preserves the mean $\Amone$ signal to within $\sim 2\%$, but the shot-noise scatter increases as $\sqrt{1/(1-f_{\rm lost})}$ as expected.  For this strongly lopsided disc ($\varepsilon_1 = 0.30$, $N = 10^5$) the mean amplitude is therefore insensitive to particle loss.  This does not hold for weakly lopsided discs sampled with fewer particles, where the shot-noise floor becomes comparable to the signal and biases $\Amone$ upwards (Sections \ref{sec:noise_floor} and \ref{sec:resolution}).

Overall, among the effects tested on synthetic discs, centring is the dominant potential systematic.  Centring offsets are not varied in the real-galaxy test of Section~\ref{sec:systematics}, where the aperture choice turns out to be the largest of the settings that are varied.  For robust centring algorithms of the type used in this work, each of the three effects is expected to contribute at the level of $\sim 10$ per cent or less, consistent with the aggregate systematic budget assessed in Section \ref{sec:systematics}.

\subsection{Mixed-component validation: coexisting bar, spiral, and lopsidedness}
\label{sec:mixed_validation}

The tests in Sections \ref{sec:validation_recovery}--\ref{sec:extended_validation} use synthetic discs with a \emph{single} injected Fourier mode.  Real galaxies, however, routinely host bars, spiral arms, and lopsidedness \emph{simultaneously}: any pipeline aimed at cosmological applications must be able to separate coexisting modes.

To characterise the pipeline's component-separability we construct a reference \emph{mixed} disc with three simultaneous perturbations,
\begin{multline}
\Sigma(R, \phi) = \Sigma_0(R) \Bigl[1 + \varepsilon_1 \cos(\phi - \phi_1) \\
+ \varepsilon_2 \cos\bigl(2(\phi - \phi_2)\bigr)
+ \varepsilon_s \cos\bigl(2(\phi - k\ln(R/\Rd))\bigr)\Bigr],
  \label{eq:mixed_disc}
\end{multline}
with injected amplitudes $\varepsilon_1 = 0.15$ (weak lopsided), $\varepsilon_2 = 0.25$ (moderate bar, offset from the lopsided phase by $45^{\circ}$), and $\varepsilon_s = 0.10$ (weak two-armed logarithmic spiral of pitch angle $p = 15^{\circ}$, with $k = \cot p \simeq 3.73$).  We generate $N = 10^5$ particles by rejection sampling from Eq. \ref{eq:mixed_disc}, running 5 random realisations.

Figure \ref{fig:fig8_mixed_validation} (Appendix \ref{app:supplementary}) summarises the outcome.  The left panel shows the surface-density map of a single realisation: the elongated bar and the outer spiral tails are visible, with a subtle lopsided offset toward positive $x$.  The right panel shows
the recovered global amplitudes $\Amone$ and $\Amtwo$ over the five realisations.  We recover $\Amone = 0.145 \pm 0.009$ (a $-3.5\%$ bias relative to $\varepsilon_1 = 0.15$) and $\Amtwo = 0.224 \pm 0.009$ (a $-11\%$ bias relative to $\varepsilon_2 = 0.25$).  The lopsided component is recovered to better than $5\%$; the small negative bias in $\Amtwo$ reflects partial destructive interference between the coherent bar phase $\phi_2$ and the radially varying phase of the logarithmic spiral, which averages down when integrated over the aperture. This is the physically expected behaviour of overlapping $m=2$ modes and can be reduced (at the cost of statistical precision) by narrower radial binning.

This mixed-component test confirms that our pipeline can reliably separate coexisting Fourier modes in the amplitude range relevant to disc lopsidedness and bar studies, with residual cross-talk at the $\sim 5$--$10\%$ level, addressing a common failure mode of naive Fourier-decomposition analyses.

\subsection{Parameter sensitivity and method robustness}
\label{sec:parameter_sensitivity}

Two of the pipeline's discretionary choices need explicit checks. The first is the coherence threshold $f_{\rm coh}^{\rm thresh}$ that we use to separate genuine bar/spiral detections from shot-noise fluctuations (Section \ref{sec:bar_length}). The second is the global amplitude-averaging convention. Recall that for each galaxy the pipeline emits three independent global amplitudes: the R\&Z95 aperture mean $\Amone^{\rm lit}$ (Eq. \ref{eq:literature_average}), the area-weighted radial integral $\Amone^{\rm area}$, and the Jog+2002 $\Sigma$-weighted integral $\Amone^{\rm Jog}$ (Eq. \ref{eq:jog_integral}). Having all three available lets us make a direct same-galaxy comparison. Figure \ref{fig:fig10_parameter_sensitivity} (Appendix \ref{app:supplementary}) presents both tests in a single panel-pair.

\emph{Threshold sensitivity} (Panel A). The bar-detection fraction tracks the coherence threshold smoothly and monotonically. It moves from $88.7\%$ at $f_{\rm coh}^{\rm thresh} = 0.3$ (permissive) through $66.7\%$ at our fiducial $f_{\rm coh}^{\rm thresh} = 0.5$ to $32.2\%$ at $f_{\rm coh}^{\rm thresh} = 0.8$ (conservative). Shifting the cut by $\pm 0.1$ around the fiducial value moves the retained fraction by $+13.4$ and $-10.2$ percentage points. We read this smooth response as evidence of no artificial bimodality at our chosen cut. Downstream statistics are much less sensitive: the median bootstrap $\Amone$ of retained galaxies is $0.097$ ($f_{\rm coh} > 0.4$), $0.097$ ($f_{\rm coh} > 0.5$) and $0.094$ ($f_{\rm coh} > 0.6$), a spread of only $\sim 3\%$. On this basis $f_{\rm coh}^{\rm thresh} = 0.5$ is a reasonable central choice.

\emph{Method robustness} (Panel~B). The three averaging methods return distributions of $\Amone$ that agree closely across the full 915-galaxy sample. Their medians match to within $\sim 1\%$ ($\Amone^{\rm lit} = 0.0914$, $\Amone^{\rm area} = 0.0924$, $\Amone^{\rm Jog} = 0.0912$). The three cumulative distributions are statistically indistinguishable: pairwise two-sample Kolmogorov--Smirnov tests give $D < 0.032$ and $p > 0.74$ in every case. Pearson correlations exceed $0.99$ across all pairs. On this basis, the R\&Z95 aperture mean we adopt in our main analysis is essentially interchangeable with either integral method. Our later comparisons to \citet{2013ApJ...772..135Z} and \citet{2022A&A...662A..53L} are not sensitive to the specific averaging convention.

Between these two tests we cover the two most obvious sources of convention-driven systematic uncertainty in the pipeline. The coherence-threshold choice moves the retained median $\Amone$ by $\sim 3\%$; the amplitude-averaging convention by $\sim 1\%$. We regard both as small compared with the physical trends we discuss in later sections.

\section{Application to Illustris TNG50}
\label{sec:tng50}

\subsection{Sample selection}
\label{sec:sample_selection}

TNG50-1 is the highest-resolution box in the IllustrisTNG suite \citep{2018MNRAS.475..676S,2018MNRAS.473.4077P,2018MNRAS.480.5113M,2018MNRAS.477.1206N}.
It runs on the moving-mesh code \textsc{AREPO} \citep{2010MNRAS.401..791S} and uses the IllustrisTNG galaxy-formation model, which includes kinetic AGN feedback \citep{2017MNRAS.465.3291W}; TNG50 itself is described by \citet{2019MNRAS.490.3234N} and \citet{2019MNRAS.490.3196P}, the latter including the first analysis of its stellar and gaseous discs. We draw all data used here from the public release \citep{2019ComAC...6....2N}.

Our starting point is snapshot 99 ($z=0$). We keep every central subhalo whose stellar mass falls in the interval $3.16\times10^9 < M_\star / \Msun < 10^{11}$. This cut yields 923 candidate subhalos. We then process the full parent list rather than a random subsample, and 915 of the 923 (99.1\%) pass our minimum disc-particle requirement. The remaining 8 are dropped because they contain fewer than 500 disc particles once our height cut $|z| < 3\kpc$ is applied. Working with the complete parent pool has two advantages over random sampling: it avoids selection bias in the amplitude distribution, and it gives us the statistics needed for a robust Kolmogorov--Smirnov comparison against observational catalogues in Section \ref{sec:comparison_zaritsky}.

The parent sample is defined by stellar mass only and is not a morphologically selected disc sample.  Where the distinction matters we therefore also use a kinematically selected subsample, referred to as \emph{filtered}: $D/T > 0.5$, where $D/T$ is the mass fraction of stars with circularity $j_z/j_{\rm circ} > 0.7$ \citep{2023MNRAS.523.5853V,2022A&A...662A..53L}, together with $\Ndisk > 10^4$ and a bootstrap-mean $\Amone < 0.5$.  It contains 446 of the 915 TNG50-1 galaxies.

\subsection{Amplitude distribution}
\label{sec:amplitude_distribution}

\begin{figure*}
  \centering
  \includegraphics[width=\textwidth]{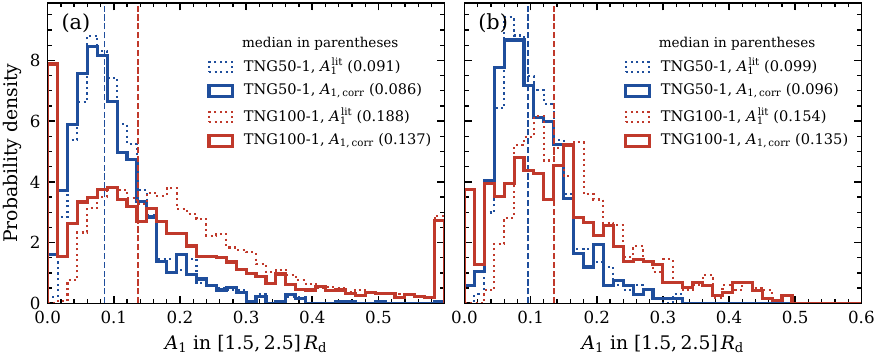}
    \caption{Distributions of $\Amone$ in the $[1.5,2.5]\,R_{\rm d}$ aperture for TNG50-1 (blue) and TNG100-1 (red), measured with identical pipeline settings.  Dotted histograms: direct aperture estimate $A_1^{\rm lit}$; solid histograms: noise-corrected $A_{1,\rm corr}$ (Section \ref{sec:noise_floor}); dashed vertical lines: medians of $A_{1,\rm corr}$.  The medians of both statistics are given in parentheses in the legends. (a) Parent samples ($N = 915$ and 6717); (b) filtered samples ($D/T > 0.5$, $\Ndisk > 10^4$, $\Amone < 0.5$; $N = 446$ and 780). The correction is small for TNG50-1 and large for TNG100-1; the peak at zero contains the galaxies whose measured $\Amone$ lies below their noise floor (11 and 5 per cent of the TNG100-1 parent and filtered samples, 1 per cent or less for TNG50-1).  Values above 0.6 (3.8 per cent of the TNG100-1 parent sample) are shown in the last bin.}
  \label{fig:fig4_A1_distribution}
\end{figure*}

Figure \ref{fig:fig4_A1_distribution} shows the distributions of the direct and noise-corrected amplitudes for both simulations.  For the full 915-galaxy TNG50-1 sample the bootstrap-mean median is $\Amone^{\rm med} = 0.097$ with a $16$--$84$ percentile range $[0.056, 0.161]$; the sample mean is $0.112 \pm 0.070$. We find 431 galaxies (47.1\%) above the $\Amone > 0.1$ threshold and 86 (9.4\%) above $\Amone > 0.2$. The distribution has a thin tail: a small number of galaxies reach $0.4 < \Amone < 0.5$, and a single object -- subhalo 668581, with $\Amone = 0.952 \pm 0.007$ -- lies well beyond that tail, roughly a factor of two above the next-highest amplitude. We return to this extreme case in Section \ref{sec:amplitude_outlier}. These are bootstrap means.  The direct aperture estimate has a median $A_1^{\rm lit} = 0.091$, and the noise-corrected amplitude of Section \ref{sec:noise_floor} a median $A_{1,\rm corr} = 0.086$; for this well-sampled simulation the three statistics differ by less than 12 per cent.

The strongly lopsided galaxies ($\Amone > 0.2$) form a natural target sample for individual case studies of accretion-driven or halo-driven $m=1$ modes. In particular, the 86 objects in this subset are the ones for which the physical mechanisms discussed in the Introduction (recent minor mergers, cosmic-web gas accretion, long-lived halo displacement) can plausibly leave a detectable imprint, and we return to them in Section \ref{sec:tng50_bardetections}.

\subsection{The single extreme outlier at $\Amone = 0.95$}
\label{sec:amplitude_outlier}

Of the 915 galaxies, exactly one -- TNG50-1 subhalo 668581 -- falls well outside the $\Amone \lesssim 0.4$ tail traced by the rest of the sample. Its bootstrap-mean amplitude is $\Amone = 0.952 \pm 0.007$, detected at $\sim 144\sigma$ against sampling variance drawn from $N_{\rm boot} = 200$ resamples. Shot noise therefore cannot explain it.

Physically, 668581 is a compact dwarf disc: $M_\star = 5.4\times10^{9}\,\Msun$, $R_d = 1.73\kpc$, $V_{\rm max} = 95\kms$. It contains $N_{\rm disk} = 8.87\times10^{4}$ particles above our selection threshold, and its shot-noise floor is $A_{1,\rm noise} = 0.02$, so the amplitude is not a small-$N$ artefact either.  Its particle set does not straddle the periodic boundary. Alongside its strong $m=1$ signal, it also hosts a coexisting $m=2$ mode at $\Amtwo = 0.30$ with pattern coherence $f_{\rm coh} = 0.91$.

Several physical scenarios are consistent with what the catalogue quantities tell us: an ongoing minor merger, a strongly displaced dark-matter halo centre, or SUBFIND segmentation of an interacting pair. None of these can be ruled out from the catalogue alone. A particle-level visual inspection combined with the SUBLINK merger-tree history would settle the question, and we defer this to future work.

We note explicitly that excluding 668581 has no material impact on any of the statistics reported in this paper. The sample median, $16$--$84$ percentile range, and the $47.1\%$ fraction above $\Amone > 0.1$ are identical to four significant figures whether the object is retained or removed. The sample mean changes only from $0.112$ to $0.111$. On this basis, we keep 668581 in the parent sample but flag it in the catalogue as a candidate for individual case study.

\subsection{Bar detections and phase coherence}
\label{sec:tng50_bardetections}

Two thirds of our sample - $66.7\%$ (610/915) - exhibit a coherent $m=2$ pattern ($f_{\rm coh} > 0.5$).  Such a pattern may be a bar or a bisymmetric spiral, so this is an upper limit on the bar fraction.  It is comparable to the total (strong plus weak) bar fraction of roughly two thirds found in near-infrared surveys of nearby discs \citep[e.g.][]{2010PASP..122.1397S,2015ApJS..217...32B}, though the different selection functions and the inclusion of spirals in our flag complicate a detailed comparison.

Note that $f_{\rm coh}$ is defined for the $m=2$ mode.  Of the 86 galaxies with $\Amone > 0.2$, 62 also have a phase-coherent $m=2$ pattern; in the rest a strong $m=1$ signal coexists with an incoherent or weak $m=2$ pattern.  This flag says nothing about the radial coherence of the $m=1$ phase itself, and we do not use it to classify lopsided galaxies.  An analogous $m=1$ coherence flag, based on the radial stability of $\phi_1(R)$, is a natural addition to the catalogue and is planned for the next release.

\subsection{Scaling with stellar mass}

The mass dependence of $\Amone$ in both simulations is shown in Fig. \ref{fig:resolution} (panels a and b).  For the TNG50-1 parent sample, splitting the sample into three logarithmically spaced mass bins reveals a clear anti-correlation. Dwarfs at $\log_{10}(M_\star/\Msun) \sim 9.7$ ($N = 442$) have a median $\Amone \approx 0.11$; intermediate-mass discs at $\sim 10.2$ ($N = 304$) sit at $\approx 0.09$; and massive discs at $\sim 10.7$ ($N = 169$) reach only $\approx 0.06$. A Spearman rank correlation on the full sample gives $\rho = -0.44$ ($p \approx 10^{-45}$); with the noise-corrected amplitude $\rho = -0.33$ ($p \approx 10^{-24}$).

Physically, the trend echoes a well-established observational finding: lower surface-density, later-type systems host stronger lopsidedness \citep{2005A&A...438..507B,2013ApJ...772..135Z}. Lopsidedness in TNG50 has also been linked to low central stellar densities and extended discs \citep{2023MNRAS.523.5853V,2023MNRAS.526.1636D}; the mass trend we find is consistent with this. We interpret this as evidence that either the physical mechanisms responsible for lopsidedness scale naturally with disc self-gravity, or that low-mass discs simply retain the imprint of recent perturbations for longer. The present data do not distinguish between these possibilities (Section \ref{sec:astrophysics}).

\subsection{Dynamical connections: rotation curve and stability}
\label{sec:dynamical}

The TNGGalaxyLab pipeline records $V_{\rm max}$ and its radial location $R_{V_{\rm max}}$ for every processed subhalo (tracer method; Section \ref{sec:algorithm}). We use these two quantities to place the recovered Fourier amplitudes in a dynamical context for all 915 galaxies.

\begin{figure*}
  \centering
  \includegraphics[width=\textwidth]{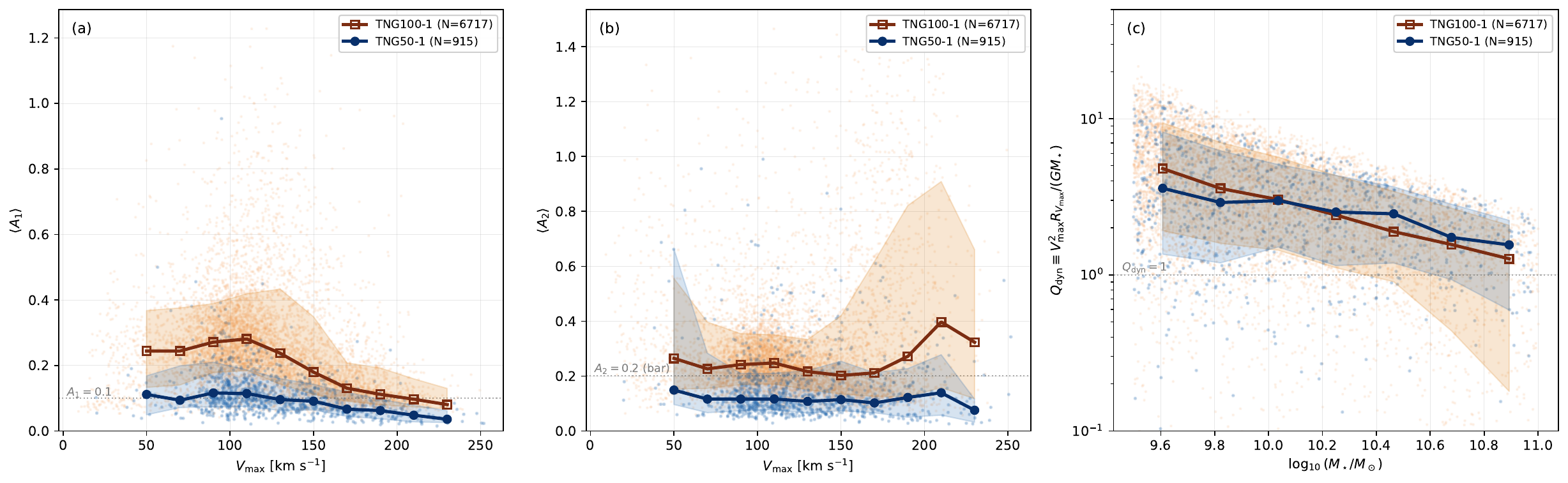}
    \caption{Dynamical connections in the raw parent samples of TNG50-1 (blue circles, $N=915$) and TNG100-1 (orange squares, $N=6717$), analysed with the identical \pipeline\ pipeline; $\langle A_m \rangle$ denotes the bootstrap-mean amplitude. (a) Lopsidedness $\langle \Amone \rangle$ vs.\ peak circular velocity $V_{\rm max}$; the dotted line marks $\Amone = 0.1$. (b) Bar/spiral amplitude $\langle \Amtwo \rangle$ vs.\ $V_{\rm max}$; the dotted line marks $\Amtwo = 0.2$. (c) Dynamical mass ratio proxy $Q_{\rm dyn} = V_{\rm max}^{2}\,R_{V_{\rm max}}/(G M_\star)$ vs.\ stellar mass; the dotted line marks $Q_{\rm dyn} = 1$. In each panel, the solid lines and shaded bands show the running median and 16--84 percentile interval in bins of the abscissa. The TNG100-1 parent sample has a larger low-$V_{\rm max}$ tail (3.5 per cent of galaxies below $50\,{\rm km\,s^{-1}}$, against 1 per cent in TNG50-1), reflecting the larger population of low-mass, marginally resolved candidate discs before quality selection.}
  \label{fig:fig9_rotation_analysis}
\end{figure*}

Figure \ref{fig:fig9_rotation_analysis}a shows $\Amone$ against $V_{\rm max}$. A clear anti-correlation runs through the data. Dwarf systems ($V_{\rm max} \sim 60$--$100\kms$) show running-median $\Amone \sim 0.11$--$0.12$; the massive rotators ($V_{\rm max} \gtrsim 200\kms$) settle at $\Amone \sim 0.04$--$0.05$. Quantitatively, we recover a Spearman correlation of $\rho = -0.39$ ($p \approx 5\times10^{-34}$).  Because $V_{\rm max}$ and $M_\star$
are tightly related, this is largely the mass trend of the previous section seen in a different variable: at fixed stellar mass the partial rank correlation between $V_{\rm max}$ and $\Amone$ is consistent with zero ($\rho_{\rm p} = 0.01$). Two physical interpretations are consistent with this. Lower-mass, later-type discs may be more susceptible to tidal interactions and asymmetric gas accretion \citep{2005A&A...438..507B,2013ApJ...772..135Z}; alternatively, their shallower potential wells may simply retain lopsided modes for longer once excited. Both are plausible, and the data alone do not distinguish between them.

Figure \ref{fig:fig9_rotation_analysis}b repeats the exercise for $\Amtwo$. In contrast to $\Amone$, the trend is essentially flat: $\rho = -0.05$, $p = 0.10$. The $\Amtwo$ median remains at $\approx 0.11$--$0.12$ across the entire mass range. This is an interesting result on its own terms. It shows that the strength of the $m=2$ pattern in TNG50 discs is largely decoupled from the total dynamical mass, in sharp contrast to the $m=1$ mode. We interpret this as evidence that $\Amtwo$ is dominated by internal disc dynamics -- bar formation, spiral driving -- rather than by the gross rotational structure that reflects the total dynamical mass.

Figure \ref{fig:fig9_rotation_analysis}c presents a dimensionless dynamical mass ratio proxy,
\begin{equation}
  Q_{\rm dyn} = \frac{V_{\rm max}^{2}\,R_{V_{\rm max}}}{G\,M_\star},
  \label{eq:Q_dyn}
\end{equation}
which approximately measures the enclosed dynamical-to-stellar mass ratio $M_{\rm dyn}(<R_{V_{\rm max}})/M_\star$ within the radius of peak rotation.  This is a compact global diagnostic that requires no velocity-dispersion measurement, and should not be confused with the classical \citet{1964ApJ...139.1217T} $Q$ parameter, with the effective multi-component stability parameter of \citet{2011MNRAS.416.1191R}, or with the Efstathiou--Lake--Negroponte bar-stability parameter \citep{1982MNRAS.199.1069E}, all of which measure distinct physical quantities.

$Q_{\rm dyn} \gg 1$ marks systems in which the total dynamical mass within $R_{V_{\rm max}}$ substantially exceeds the stellar mass (i.e.\ dominated by dark matter and gas); $Q_{\rm dyn} \sim 1$ identifies systems where the stellar component contributes appreciably to the enclosed dynamical mass; $Q_{\rm dyn} < 1$ indicates stellar-dominated potentials within this aperture.  Our TNG50 sample spans more than a decade in $Q_{\rm dyn}$ with a median value $Q_{\rm dyn}^{\rm med} = 2.6$ and a running-median trend that decreases monotonically with stellar mass (Spearman $\rho = -0.36$, $p < 10^{-25}$), from $Q_{\rm dyn} \sim 3.5$ at $\log_{10}(M_\star/\Msun) \sim 9.5$ to $\sim 1.5$ at $\log_{10}(M_\star/\Msun) \sim 10.8$.  A significant positive Spearman correlation between $Q_{\rm dyn}$ and $\Amone$ is detected ($\rho = +0.35$, $p < 10^{-25}$).  Part of this reflects the common dependence of both quantities on stellar mass, but not all of it: at fixed $M_\star$ the partial rank correlation is $\rho_{\rm p} = +0.23$ for the bootstrap-mean amplitude and $+0.18$ for $A_{1,\rm corr}$.  In our TNG50 sample, systems with the smallest $Q_{\rm dyn}$ (stellar-dominated regime) correspond preferentially to the massive end of the population.  In contrast, the highest-$Q_{\rm dyn}$ systems (dark-matter dominated within $R_{V_{\rm max}}$) coincide with the dwarf-mass, most lopsided galaxies.

In short, strong lopsidedness in TNG50-1 is found preferentially in low-mass systems whose inner mass budget is dominated by dark matter and gas.  This is the regime in which lopsided modes are expected to be excited easily and to persist \citep{2009PhR...471...75J}, although the present data do not identify the mechanism.

\section{Comparison with simulations and observations}
\label{sec:comparison}

\subsection{Comparison with the S$^{4}$G sample of Zaritsky et al.\ (2013)}
\label{sec:comparison_zaritsky}

\citet{2013ApJ...772..135Z} apply an $m=1$ Fourier decomposition to a sample of 167 nearby galaxies observed at $3.6\,\mu$m as part of the Spitzer Survey of Stellar Structure in Galaxies \citep[S$^{4}$G;][]{2010PASP..122.1397S,2015ApJS..217...32B} -- a volume-limited imaging survey of 2352 nearby galaxies with uniform photometry \citep{2015ApJS..219....3M} and morphological classification \citep{2015ApJS..217...32B} -- using the identical inner radial aperture $[1.5, 2.5]\,\Rd$ as our pipeline.  Our stellar-mass matching adopts the $3.6\,\mu$m mass-to-light calibration of \citet{2014ApJ...788..144M} with the solar absolute magnitude of \citet{2018ApJS..236...47W}.  This allows a methodologically consistent comparison between the two samples.

\begin{figure*}
  \centering
  \includegraphics[width=\textwidth]{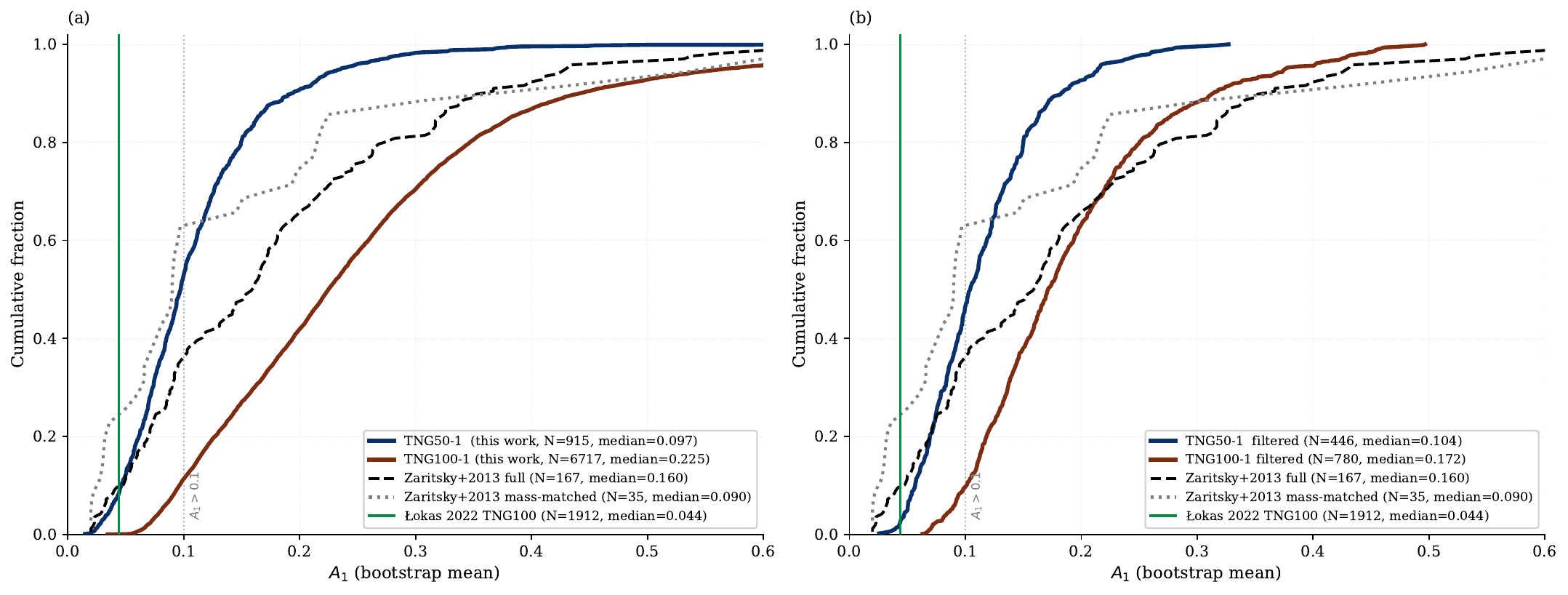}
   \caption{Cumulative distributions of the bootstrap-mean $\Amone$. (a) Parent samples: TNG50-1 (blue, $N=915$) and TNG100-1 (red, $N=6717$) from this work, the full S$^{4}$G sample of \citet{2013ApJ...772..135Z} (dashed black, $N=167$) and its mass-matched subsample with $10^{9.5} < M_\star/\Msun < 10^{11}$ (dotted grey, $N=35$).  The green vertical line marks the median $\Amone = 0.044$ reported for TNG100 by \citet[][L22]{2022A&A...662A..53L}, measured in a different aperture ($[1,2]\,r_{1/2}$); only the median is published.  The dotted vertical line marks $\Amone = 0.1$.  (b) As (a), with the TNG samples restricted to the filtered subsamples ($D/T > 0.5$, $\Ndisk > 10^4$, $\Amone < 0.5$).  Two-sample KS tests for the parent samples are given in Section \ref{sec:comparison_zaritsky}; for the filtered samples, TNG50-1 gives $D = 0.346$ ($p = 1.9\times10^{-13}$) against the full and $D = 0.234$ ($p = 0.048$) against the mass-matched observed sample, and TNG100-1 gives $D = 0.270$ ($p = 2.6\times10^{-9}$) and $D = 0.541$ ($p = 1.1\times10^{-9}$).}
  \label{fig:fig6_obs_comparison}
\end{figure*}

Figure \ref{fig:fig6_obs_comparison} compares the cumulative distributions of $\Amone$.  Two comparisons are informative.

\emph{Full sample.}  Using the full 167-galaxy Zaritsky sample, a two-sample Kolmogorov--Smirnov test yields $D = 0.338$, $p < 10^{-14}$; the observed median (0.160) is $1.64\times$ larger than our TNG50 median (0.097).  However, the Zaritsky sample spans a wider morphology range and includes numerous
low-mass galaxies with median $\log_{10}(M_\star / \Msun) \simeq 9.0$, well below our TNG50 selection floor of $10^{9.5}$.

\emph{Mass-matched subsample.}  Restricting the Zaritsky sample to the same stellar-mass range as our TNG50 selection ($10^{9.5} < M_\star / \Msun < 10^{11}$; 35 galaxies) yields a subsample median $\Amone = 0.090$.  The corresponding Kolmogorov--Smirnov test yields $D = 0.214$, $p = 0.079$: the null hypothesis of a common parent distribution cannot be rejected at the 95\% confidence level.  The observed-to-simulated median ratio of $0.93$ (TNG50 slightly higher than the mass-matched observations) indicates that, within the statistical uncertainty of the small observational subsample, the two distributions are consistent.

Within the statistical power of a 35-galaxy comparison sample, the TNG50-1 parent sample is therefore consistent with the mass-matched S$^{4}$G sample of \citet{2013ApJ...772..135Z}, and most of the apparent offset with respect to the full Zaritsky sample is due to its low-mass tail.  The agreement is not unconditional: for the filtered TNG50-1 subsample the KS test gives $D = 0.234$ and $p = 0.048$, a marginal rejection at the 95 per cent level, and a sample of 35 galaxies cannot exclude differences of $\sim$30 per cent in the median.  TNG100-1 is inconsistent with both observed samples (mass-matched: $D = 0.526$, $p \approx 2\times10^{-9}$; full: $D = 0.255$, $p \approx 8\times10^{-10}$; Fig. \ref{fig:fig6_obs_comparison}). These tests use the bootstrap-mean amplitudes plotted in the figure. For TNG50-1 the choice of statistic makes little difference (Section \ref{sec:amplitude_distribution}); for TNG100-1 the noise-corrected median is 0.137 rather than 0.225, which reduces but does not remove the excess with respect to the mass-matched observations.

\emph{Consistency checks with alternative statistical measures.} To ensure that our conclusions do not depend on the choice of the Kolmogorov--Smirnov statistic alone, we repeat both comparisons with three additional measures: the Mann--Whitney $U$ test \citep[rank-based location shift;][]{10.1214/aoms/1177730491}, the Anderson--Darling $k$-sample test \citep[weighted sensitivity to distribution tails;][]{Scholz01091987}, and Cliff's $\delta$ \citep[non-parametric effect size;][]{Cliff1993DominanceSO}.  For the TNG50-vs.-mass-matched comparison, all four tests are consistent with the KS conclusion of statistical indistinguishability: Mann-Whitney $U = 18310$ ($p = 0.15$), Anderson-Darling $T = 0.23$ ($p = 0.25$), and Cliff's $\delta = +0.14$, corresponding to a ``negligible'' effect on the scale of \citet{Romano2006}.  For the TNG50-vs.-full-sample comparison, all four tests confirm a highly significant difference, with Mann--Whitney $p = 10^{-23}$, Anderson--Darling $T = 66.8$, and a ``large'' effect size ($\delta = -0.49$; Cohen's $d = -0.85$, \citealt{cohen1988statistical}). A tabulated summary of all statistics is provided in the supplementary catalogue release (\texttt{additional\_statistics.md}).  These checks confirm that the KS result for TNG50-1 is not specific to the choice of test.

\subsection{TNG100-1 with the same pipeline, and the comparison with \L{}okas (2022)}
\label{sec:comparison_lokas}

\citet{2022A&A...662A..53L} analysed 1912 disc galaxies from TNG100 at $z=0$ with an $m=1$ Fourier decomposition averaged over $[1, 2]\,r_{1/2}$, and found a mean $\langle \Amone \rangle = 0.051$, a median of 0.044 and only 8.4 per cent of galaxies with $\Amone > 0.1$.  Set against our TNG50-1 median of 0.097, this would suggest that TNG50 discs are about twice as lopsided as TNG100 discs, i.e.\ that lopsidedness is under-resolved in TNG100.  That comparison, however, mixes two different measurements: the aperture, the centring and disc selection, and possibly the mode normalisation all differ between the two studies. The aperture alone can change the result substantially: \citet{2023MNRAS.526.1636D} find that the lopsided fraction of TNG50 discs is 10-30 \% in radial ranges similar to those used in observations, but rises to 58 \% when $\Amone$ is measured at larger radii.

To compare the two simulations directly we ran \pipeline\ with identical settings on all 6743 central galaxies of TNG100-1 at $z=0$ with $3.16\times10^9 < M_\star/\Msun < 10^{11}$; 6717 were processed successfully and 26 failed the disc-particle or centring requirements.  The results are summarised in Table \ref{tab:comparison_new}, for three statistics: the bootstrap mean used in earlier versions of the catalogue, the direct aperture estimate $A_1^{\rm lit}$, and the noise-corrected $A_{1,\rm corr}$.
\begin{table*}
  \centering
  \caption{Median $\Amone$ in simulations and observations.  All ``this work'' rows use the $[1.5, 2.5]\,R_{\rm d}$ aperture and identical pipeline settings.  The S$^{4}$G values of \citet{2013ApJ...772..135Z} are measured in the same aperture on 3.6\,$\mu$m images and are listed in the $A_1^{\rm lit}$ column.  The value of \citet{2022A&A...662A..53L} refers to $[1,2]\,r_{1/2}$ and is not directly comparable. ``Filtered'': $D/T > 0.5$, $\Ndisk > 10^4$ and bootstrap-mean $\Amone < 0.5$.}
\label{tab:comparison_new}
\begin{tabular}{lr ccc cc}
\toprule
& & \multicolumn{3}{c}{Median $\Amone$}
& \multicolumn{2}{c}{Fraction with $\Amone > 0.1$} \\
\cmidrule(lr){3-5} \cmidrule(lr){6-7}
Sample & $N$ & bootstrap mean & $A_1^{\rm lit}$ & $A_{1,\rm corr}$
& bootstrap mean & $A_{1,\rm corr}$ \\
\midrule
\multicolumn{7}{l}{\emph{This work}} \\
TNG50-1, parent            &  915 & 0.097 & 0.091 & 0.086 & 47.1\% & 39.8\% \\
TNG100-1, parent           & 6717 & 0.225 & 0.188 & 0.137 & 88.7\% & 62.3\% \\
TNG50-1, filtered          &  446 & 0.104 & 0.099 & 0.096 & 53.6\% & 46.6\% \\
TNG100-1, filtered         &  780 & 0.172 & 0.154 & 0.135 & 90.4\% & 64.1\% \\
\midrule
\multicolumn{7}{l}{\emph{Literature}} \\
S$^{4}$G, full             &  167 & ---   & 0.160 & ---   & 64.1\% & ---    \\
S$^{4}$G, mass-matched     &   35 & ---   & 0.090 & ---   & 37.1\% & ---    \\
\L{}okas (2022)            & 1912 & ---   & 0.044 & ---   & 8.4\%  & ---    \\
\bottomrule
\end{tabular}
\end{table*}

With the measurement held fixed, the ordering of the two simulations is the opposite of the one suggested by the cross-study comparison.  TNG100-1 is \emph{more} lopsided than TNG50-1 in every statistic and in both samples, and the difference is highly significant (KS $p < 10^{-17}$ in all cases).  Its size, however, depends strongly on the statistic.  For the parent samples the TNG100/TNG50 ratio of medians is 2.31 for the bootstrap mean, 2.06 for $A_1^{\rm lit}$ and 1.59 for $A_{1,\rm corr}$; for the filtered samples it is 1.64, 1.55 and 1.41.  The bootstrap and shot-noise biases of Section \ref{sec:algorithm} thus account for roughly half of the raw difference, and Section \ref{sec:resolution} examines what remains.
\begin{table*}
  \centering
  \caption{TNG50-1/TNG100-1 ratio of median $\Amone$ for successively stricter selections, applied identically to both simulations, for the bootstrap mean, the direct aperture estimate and the noise-corrected amplitude.  The last column is the KS $p$-value for $A_{1,\rm corr}$.}
  \label{tab:filter_progression}
  \begin{tabular}{l rr ccc r}
    \toprule
    & \multicolumn{2}{c}{$N$}
      & \multicolumn{3}{c}{TNG50/TNG100, median $\Amone$}
      & KS $p$ \\
    \cmidrule(lr){2-3} \cmidrule(lr){4-6} \cmidrule(lr){7-7}
    Selection & TNG50 & TNG100
      & bootstrap mean & $A_1^{\rm lit}$ & $A_{1,\rm corr}$
      & ($A_{1,\rm corr}$) \\
    \midrule
    Parent sample                         & 915 & 6717 & 0.433 & 0.485 & 0.628 & $4\times10^{-64}$ \\
    $\Amone < 0.5$                        & 914 & 6234 & 0.454 & 0.512 & 0.684 & $6\times10^{-47}$ \\
    $D/T > 0.4$                           & 629 & 2860 & 0.456 & 0.506 & 0.624 & $5\times10^{-48}$ \\
    $D/T > 0.5$                           & 446 & 1154 & 0.507 & 0.555 & 0.663 & $3\times10^{-28}$ \\
    $D/T > 0.5$, $\Ndisk > 10^4$          & 446 &  800 & 0.597 & 0.627 & 0.695 & $1\times10^{-20}$ \\
    $D/T > 0.5$, $\Ndisk > 10^4$, $\Amone < 0.5$ & 446 & 780 & 0.608 & 0.644 & 0.708 & $2\times10^{-18}$ \\
    \bottomrule
  \end{tabular}
\end{table*}
Our TNG100-1 medians are three to five times larger than the value of \citet{2022A&A...662A..53L} for the same simulation.  We have not re-measured her sample in her aperture, so we cannot say how this factor divides between aperture, normalisation, sample selection and noise.  What the comparison does show is that the low TNG100 value should not be interpreted as evidence that TNG100 under-produces lopsidedness relative to TNG50: measured in the same way, it produces more.

Our results complement the TNG50 studies of \citet{2023MNRAS.523.5853V}, \citet{2023MNRAS.526.1636D}, \citet{2025A&A...693A..63D} and \citet{2025A&A...693A..64F}, which address the physical drivers and evolution of lopsidedness within a single simulation; here the focus is on how the measured amplitude depends on resolution and on the estimator.

\subsection{Measurement bias versus resolution dependence}
\label{sec:resolution}
\label{sec:methodology_new}

We now separate the TNG100/TNG50 difference into a part caused by the measurement and a part intrinsic to the simulated galaxies. The tests below use the filtered samples and compare the two simulations at fixed stellar mass, because the filtered TNG100-1 sample contains almost no galaxies below $10^{10}\,\Msun$ (the $\Ndisk > 10^4$ requirement removes them) while half of the filtered TNG50-1 sample lies there.

\paragraph*{Filter progression.} Table \ref{tab:filter_progression} follows the medians through successively stricter selections.  Restricting to kinematic discs reduces the offset, mainly by removing the poorly sampled low-mass end of TNG100-1, but at every step the noise correction reduces it further.  In the strictest selection the TNG50/TNG100 ratio is 0.61 for the bootstrap mean and 0.71 for $A_{1,\rm corr}$.

\paragraph*{Downsampling TNG50-1.} The most direct test of the noise correction is to degrade TNG50-1 to the resolution of TNG100-1 and see whether the correction undoes the degradation.  For each filtered TNG50-1 galaxy we kept a random fraction $f = m_{\rm b}^{\rm TNG50}/m_{\rm b}^{\rm TNG100}= 8.5\times10^4/1.4\times10^6 \approx 1/16$ of the star particles, multiplied their masses by $1/f$, repeated the centring, alignment and Fourier analysis from scratch, and took the median over three random realisations.  The aperture was held at the catalogue $R_{\rm d}$, so that only the particle number changes.  These tests use a stand-alone re-implementation of Steps 1--6 of Fig. \ref{fig:pipeline_flowchart}, distributed with the code; it reproduces the catalogue $A_1^{\rm lit}$ of TNG100-1 exactly (median ratio 1.000) and that of TNG50-1 to 6 per cent (Spearman $\rho = 0.98$), and all samples in the test are measured with it.

\begin{figure*}
  \centering
  \includegraphics[width=\textwidth]{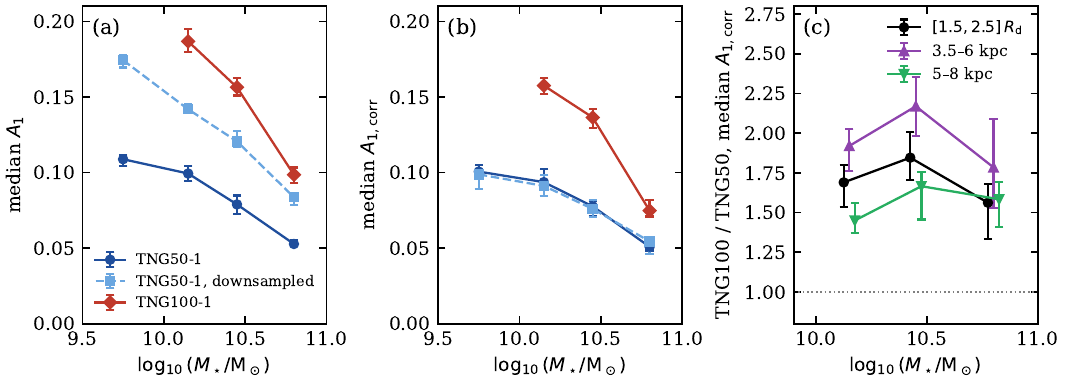}
    \caption{Decomposition of the TNG100/TNG50 difference for the filtered samples.  (a) Median uncorrected and (b) median noise-corrected $\Amone$ in bins of stellar mass for TNG50-1 (dark blue), TNG50-1 randomly downsampled to the TNG100-1 particle mass (light blue, dashed) and TNG100-1 (red), all measured with the same code in the $[1.5,2.5]\,R_{\rm d}$ aperture.  Downsampling inflates the uncorrected amplitude by 43--60 per cent but leaves $A_{1,\rm corr}$ unchanged to within 2--7 per cent; TNG100-1 lies above both.  (c) Ratio of the TNG100-1 to the TNG50-1 median $A_{1,\rm corr}$ from the pipeline catalogues, in the $[1.5,2.5]\,R_{\rm d}$ aperture and in two fixed physical apertures; the dotted line marks a ratio of unity. Error bars are 16--84 per cent bootstrap intervals on the medians.}
  \label{fig:resolution}
\end{figure*}

Table \ref{tab:downsampling} and Fig. \ref{fig:resolution} give the result.  Downsampling raises the uncorrected median $\Amone$ of TNG50-1 by 43--60 per cent in every mass bin, purely through the loss of particles.  After correction the downsampled values agree with the full-resolution ones to within 2--7 per cent.  The correction therefore works at the particle numbers of TNG100-1.  In the two fixed physical apertures introduced below the agreement is slightly worse, with the correction over-subtracting by up to 13 per cent at $N_{\rm eff} \approx 300$, which we adopt as its systematic uncertainty.  The galaxy-by-galaxy scatter is large: the 16--84 per cent range of the ratio of downsampled to full-resolution $A_{1,\rm corr}$ is 0.17--1.35, and 19 per cent of downsampled realisations fall entirely below their noise floor ($A_{1,\rm corr} = 0$).  The correction is therefore reliable for population medians, which is how we use it, but not for individual low-amplitude galaxies at TNG100-1 resolution.

\begin{table*}
  \centering
  \caption{Downsampling test, filtered samples, aperture $[1.5,2.5]\,R_{\rm d}$.  For each stellar-mass bin we list the number of galaxies, the median uncorrected $\Amone$, the median $A_{1,\rm corr}$ and the median effective particle number per aperture bin, for TNG50-1 at full resolution, TNG50-1 downsampled to the TNG100-1 particle mass, and TNG100-1.}
  \label{tab:downsampling}
  \begin{tabular}{l rccr rccr rccr}
    \toprule
    & \multicolumn{4}{c}{TNG50-1}
      & \multicolumn{4}{c}{TNG50-1 downsampled}
      & \multicolumn{4}{c}{TNG100-1} \\
    \cmidrule(lr){2-5} \cmidrule(lr){6-9} \cmidrule(lr){10-13}
    $\log_{10}(M_\star/\Msun)$
      & $N$ & $\Amone$ & $A_{1,\rm corr}$ & $N_{\rm eff}$
      & $N$ & $\Amone$ & $A_{1,\rm corr}$ & $N_{\rm eff}$
      & $N$ & $\Amone$ & $A_{1,\rm corr}$ & $N_{\rm eff}$ \\
    \midrule
    9.5--10.0  & 227 & 0.109 & 0.101 & 2523  & 227 & 0.174 & 0.099 & 151 & --- & ---   & ---   & --- \\
    10.0--10.3 & 111 & 0.099 & 0.094 & 4900  & 111 & 0.142 & 0.091 & 290 & 290 & 0.187 & 0.158 & 396 \\
    10.3--10.6 &  73 & 0.079 & 0.078 & 7558  &  73 & 0.121 & 0.076 & 451 & 332 & 0.156 & 0.136 & 621 \\
    10.6--11.0 &  34 & 0.053 & 0.051 & 15053 &  34 & 0.084 & 0.055 & 920 & 158 & 0.099 & 0.075 & 951 \\
    \bottomrule
  \end{tabular}
\end{table*}

The same table shows what remains.  In the two upper mass bins the downsampled TNG50-1 discs have nearly the same aperture particle numbers as TNG100-1 ($N_{\rm eff} = 451$ vs 621 and 920 vs 951), so the uncorrected amplitudes can be compared without any model for the noise: TNG100-1 is still higher, by factors of 1.2--1.3.  After correction the excess at fixed mass is 1.4--1.8 in this test (Fig. \ref{fig:resolution}b) and 1.6--1.9 in the pipeline
catalogues (Fig. \ref{fig:resolution}c).  Shot noise therefore explains between a half and two thirds of the raw TNG100/TNG50 difference in these bins, but not all of it.  The high-$\Amone$ tail of the TNG100-1 parent sample is also not a noise effect: 483 galaxies have a bootstrap-mean $\Amone > 0.5$ and 418 still have $A_{1,\rm corr} > 0.5$.  Only 23 of these 418 have $D/T > 0.5$, and their median $\Ndisk$ is $7\times10^3$: the tail consists mostly of poorly resolved, kinematically hot or disturbed systems.

\paragraph*{Aperture definition and disc size.} At fixed stellar mass the filtered TNG100-1 discs have scale lengths 17--29 per cent smaller than those of TNG50-1 (median $R_{\rm d} = 2.4$--2.7 against 3.2--3.3\,kpc), so the $[1.5,2.5]\,R_{\rm d}$ aperture samples smaller physical radii in TNG100-1.  To remove this difference we repeated the measurement in two fixed physical apertures.  The excess persists in both (Fig. \ref{fig:resolution}c): the TNG100/TNG50 ratio of median $A_{1,\rm corr}$ is 1.8--2.2 in the 3.5--6\,kpc aperture and 1.5--1.7 in the 5--8\,kpc aperture.  The two simulations also differ in radial profile: in TNG50-1 the corrected amplitude rises from the inner to the outer aperture, as in observed discs, whereas in TNG100-1 it is nearly flat, so the excess is concentrated at 3.5--6\,kpc.

\paragraph*{Disc kinematics.} The filtered TNG100-1 discs are slightly hotter than the TNG50-1 discs at the same mass (median $\kappa_{\rm rot} = 0.70$--0.71 against 0.71--0.76; $D/T = 0.54$--0.56 against 0.58--0.64), as expected from the larger dark-matter particle mass of TNG100 \citep[e.g.][]{2019MNRAS.488L.123L}.  To test whether this explains the excess we fitted $\log A_1 = a + b\,[\mathrm{TNG100}] + c\,(\log M_\star - 10.3) + d\,(\kappa_{\rm rot} - \tilde\kappa) + e\,(D/T - \widetilde{D/T})$ to the combined filtered samples, where $[\mathrm{TNG100}]$ is an indicator variable and the tildes denote sample medians, and read the resolution effect at fixed mass and kinematics from $10^{b}$. With mass alone, $10^{b} = 1.73$ (68 per cent bootstrap range 1.66--1.80) for $A_{1,\rm corr}$ in $[1.5,2.5]\,R_{\rm d}$; adding $\kappa_{\rm rot}$ gives 1.84, and adding both $\kappa_{\rm rot}$ and $D/T$ gives 1.71.  The corresponding values are 2.13, 2.33 and 2.15 in the 3.5--6\,kpc aperture, and 1.63, 1.67 and 1.58 in the 5--8\,kpc aperture.  For the uncorrected amplitudes of TNG100-1 and of the downsampled TNG50-1, which have matched particle numbers, the mass-only value is 1.32 and does not change appreciably when the kinematic terms are added.  Within each simulation, discs with higher $\kappa_{\rm rot}$ are slightly \emph{more} lopsided at fixed mass, so controlling for the hotter TNG100-1 discs increases rather than reduces the excess.  ($\kappa_{\rm rot}$ and $D/T$ are strongly correlated, and their individual coefficients are not separately meaningful.)

\paragraph*{Interpretation.} The difference between TNG100-1 and TNG50-1 thus has three parts: the bootstrap bias (11--19 per cent in TNG100-1, 5--7 per cent in TNG50-1), the shot-noise bias (between a half and two thirds of the remaining raw difference at fixed mass), and a residual factor of 1.5--1.9 at fixed mass that is independent of the particle number in the aperture, of the aperture definition and of global disc kinematics, and is largest at 3.5--6\,kpc.  The residual is not produced by the measurement steps tested here, but we cannot assign it to a single cause. The two simulations differ in box size as well as in resolution, so their galaxy populations differ in environment and assembly history even at fixed stellar mass, and we have not matched them in halo mass, environment or merger history.  Among numerical effects, two could produce such an excess.  The stellar softening length at $z=0$ is $0.74\kpc$ in TNG100-1 and $0.29\kpc$ in TNG50-1, which changes the inner potential that a lopsided perturbation responds to. The dark-matter particle mass is about 16 times larger in TNG100-1, and discreteness noise in a halo can excite weakly damped $m=1$ modes \citep{1998MNRAS.297..101W}.  Our data cannot distinguish between these possibilities; simulations in which the two resolutions are varied separately would.  If the residual is numerical, as its concentration in the inner disc suggests, TNG50-1 is the better converged of the two runs for lopsidedness studies, and TNG100-1 amplitudes, even after the noise correction, are best treated as upper limits on the converged values.

\subsection{Mock-observation comparison}
\label{sec:mock_observations}

The mock-observation cascade described in this subsection was developed for TNG50-1.  A matched cascade for TNG100-1 would have to deal with the much larger shot-noise floor of that run (Section \ref{sec:resolution}) and is left for future work.

A residual concern with any comparison between TNG50 and the \citet{2013ApJ...772..135Z} S$^{4}$G sample is that the two are technically different measurements. TNG50 gives us particle positions and masses in three dimensions with essentially perfect signal-to-noise. S$^{4}$G, by contrast, is a photometric survey and comes with point-spread function (PSF) broadening, sky-background noise, magnitude limits, and unknown inclinations. To test whether this observational degradation contributes materially to the residual TNG50--observation offset, we build a suite of \emph{mock observations} of a 100-galaxy TNG50 subsample and re-run the identical Fourier pipeline on the degraded data.

Each mock observation applies the following degradation cascade: (i) random inclination sampled isotropically in $\cos i$ over $0^{\circ} \le i \le 60^{\circ}$; (ii) random distance sampled uniformly in $10 \le d/\mathrm{Mpc} \le 40$, spanning the S$^{4}$G distance range; (iii) projection of stellar particles onto a $256 \times 256$ pixel mass map with a $\pm 20\kpc$ field of view; (iv) convolution with a Gaussian PSF of full width $2''$ FWHM (matching Spitzer IRAC ch1, the S$^{4}$G $3.6\mu\mathrm{m}$ imaging band), scaled to the target distance; (v) addition of Gaussian noise with peak SNR $\sim 30$ per pixel, matching typical S$^{4}$G surface-brightness sensitivity in the disc; and (vi) detection thresholding at $3\sigma$ above the noise floor, before converting detected pixels back to pseudo-particles for the Fourier pipeline. Each galaxy is observed at three independent random combinations of inclination and distance to sample the observational parameter space.

Figure \ref{fig:fig11_mock_comparison} (Appendix \ref{app:supplementary}) summarises the outcome for 300 mock observations of 100 subhalos. Panel A compares the recovered $\Amone$ for each mock against the raw TNG50 measurement. The two are positively correlated but with substantial scatter and a systematic upward offset: the mock median $\Amone = 0.122$ exceeds the raw median of $0.106$ for the same subsample, a positive bias of $\sim 15\%$.
Panel B plots the recovered bias $\Delta \Amone = \Amone^{\rm mock} - \Amone^{\rm raw}$ against inclination; the running-median bias is approximately flat at $\sim +0.015$ across the sampled inclination range $0^\circ \le i \le 60^\circ$, consistent with the expectation from the extended-validation analysis in Section \ref{sec:extended_validation} that inclination alone contributes at the $\lesssim 3\%$ level for $i \lesssim 30^\circ$. Panel C shows the cumulative distributions of the raw and mock $\Amone$ samples: the mock distribution is shifted rightward at intermediate amplitudes, and
statistically distinct from the raw distribution at high significance ($D_{\rm KS} = 0.235$, $p = 1.6 \times 10^{-7}$).

The upward shift is what one expects from Section \ref{sec:noise_floor}. Pixel noise, PSF smoothing of a finite number of particles, and the detection threshold all reduce the effective number of independent resolution elements in the aperture, and any loss of information raises a positive-definite amplitude.  The detection threshold also removes the faint outer disc preferentially, which changes the region over which the aperture average is taken. We have not separated these contributions.

This has a concrete consequence for comparing simulations with observations, and in particular for the direct raw-particle comparison of Section \ref{sec:comparison_zaritsky}. Applied without an inclination and detection-threshold correction, our mock cascade drives the TNG50 median from $0.106$ (raw, on the same 100-galaxy subsample) up to $0.122$, exceeding the mass-matched \citet{2013ApJ...772..135Z} median of $0.090$ by roughly $36\%$. This overshoot is larger than the $\sim 8\%$ residual seen in the direct raw-vs.-observation comparison, indicating that the mock degradation, as currently implemented, over-estimates the observational enhancement of $\Amone$. Two effects are likely responsible: the mock 100-galaxy subsample is not itself mass-matched to Zaritsky+2013 (its raw median $0.106$ already exceeds the full-sample median of $0.097$), and the applied detection threshold is a conservative approximation of the actual S$^{4}$G sky-subtraction and $3\sigma$ ellipse-fitting pipeline. We therefore treat the direct raw-particle comparison of Section \ref{sec:comparison_zaritsky} as our fair simulation-vs.-observation reference, and interpret the mock cascade primarily as
a demonstration that plausible observational systematics push the recovered $\Amone$ upward rather than downward. Quantifying the enhancement precisely would require a mock pipeline matching the S$^{4}$G ellipse-fitting and aperture-selection procedures on a mass-matched TNG50 subsample, which we defer to future work.

\section{Systematic uncertainties}
\label{sec:systematics}

The uncertainties quoted in earlier sections are 1$\sigma$ bootstrap errors capturing shot noise; here we address the systematic uncertainties introduced by discretionary pipeline choices.  Because a comprehensive re-analysis of the full sample with each alternative setting is computationally expensive and beyond the immediate scope of this methodology paper, we estimate the systematic budget by re-running the pipeline with alternative settings on a random subsample of 30 TNG50-1 galaxies, and compare the result with our synthetic-galaxy validation (Section \ref{sec:validation}) and with published sensitivity studies for comparable Fourier pipelines \citep{2003MNRAS.341.1179A,2005A&A...434..109A,2013ApJ...772..135Z}.

Table \ref{tab:systematics} summarises the fractional effect on $\Amone$ of five alternative settings, measured on the 30 real galaxies, and of two further effects (centring and inclination) estimated from synthetic discs only.

\begin{table}
  \centering
  \caption{Systematic uncertainty in $\langle \Amone \rangle$ measured on a random sample of 30 TNG50 galaxies by re-running the fiducial pipeline with each alternative pipeline setting. Values are the median $|\Delta \Amone / \Amone|$ over the 30 galaxies, with the 16--84th-percentile range in brackets. Extended-validation synthetic estimates from Section \ref{sec:extended_validation} are also shown for comparison.}
  \label{tab:systematics}
  \begin{tabular}{lcc}
    \toprule
    Systematic       & Real (30-galaxy) & Synthetic \\
                     & median [16-84\%] & (Sec. \ref{sec:extended_validation})\\
    \midrule
    Aperture: $[1.5,2.5]$ vs.\ $[1.0,3.0]\,\Rd$  & 20.5\% [7--54]  & $\sim$5\% \\
    Radial binning: 20 bins                     & 5.3\% [1--15]   & $\sim$1\% \\
    Radial binning: 80 bins                     & 5.7\% [2--20]   & $\sim$1\% \\
    Height cut: $|z|<2\kpc$                     & 7.1\% [3--11]   & --- \\
    Height cut: $|z|<5\kpc$                     & 3.9\% [1--8]    & --- \\
    Centring: 0.1 kpc offset                    & ---              & $\sim$10\% \\
    Inclination: 30$^\circ$ vs.\ face-on         & ---              & $\sim$3\% \\
    \midrule
    Combined (quadrature)$^a$                   & \textbf{22.4\%}  & $\sim$12\% \\
    \bottomrule
  \end{tabular}
  \\[2pt]
  \raggedright\footnotesize
  $^a$ Quadrature sum uses the larger of each paired variant (aperture $20.5\%$, radial-binning $5.7\%$ from the 80-bin choice, height-cut $7.1\%$ from the $|z|<2\kpc$ choice), avoiding double-counting of the two-sided sensitivity tests.
\end{table}

The extended synthetic-galaxy validation of Section \ref{sec:extended_validation} identified centring accuracy as the dominant systematic among the effects tested there.  Here we extend the assessment by re-running the pipeline with each of five alternative discretionary choices on a random sample of 30 TNG50 galaxies (Table \ref{tab:systematics}, middle column).  Two features of the real-data budget stand out.

First, the aperture-choice systematic emerges as the \emph{single largest} contributor at $\sim 20\%$ median (16-84th percentiles $[7, 54]\%$)-- substantially larger than the $\sim 5\%$ inferred from a single, idealised synthetic galaxy.  The reason is that real cosmological discs display substantial radial variation in $\Amone(R)$: switching between the fiducial $[1.5, 2.5]\,\Rd$ and the alternative $[1.0, 3.0]\,\Rd$ aperture samples different disc regions with materially different lopsidedness levels.  This finding is qualitatively consistent with \citet{2013ApJ...772..135Z} but larger in magnitude for our TNG50 sample, likely because a cosmological population has broader intrinsic radial-profile diversity.

Second, the radial-binning systematic ($\sim 5\%$) is also larger in real galaxies than in the synthetic tests, because coarser binning smooths over intrinsic radial gradients that are absent in the (radially uniform by construction) synthetic discs. Height-cut variation contributes $\sim 4$--$7\%$, sub-dominant to aperture and comparable to binning.

The quadrature sum of the median shifts, taking the larger of each paired variant to avoid double-counting the two-sided sensitivity tests, is $\sim 22\%$ (Table \ref{tab:systematics}, bottom row) -- a factor of two larger than the synthetic-galaxy estimate. This more realistic figure indicates that the residual $\sim 8\%$ median offset between TNG50 and the mass-matched Zaritsky et al.\ (2013) sample (Section \ref{sec:comparison_zaritsky}) lies within our estimated systematic budget.  This budget is based on 30 galaxies and five alternative settings, and should be read as an order of magnitude rather than a strict bound; the agreement is consistent with, but not explained by, the systematics.  The TNG100/TNG50 difference is different in kind: it is a comparison between two simulations measured with identical settings, so the discretionary choices in Table \ref{tab:systematics} largely cancel.  The relevant systematic there is the resolution-dependent noise bias, which is not part of this budget and is treated separately in Section \ref{sec:resolution}.

\section{Astrophysical implications}
\label{sec:astrophysics}

The TNG50-1 catalogue also shows three astrophysical trends.  We discuss them briefly; a proper test of the mechanisms requires the merger histories and is beyond the scope of this paper.

\subsection{The lopsidedness--rotation anti-correlation}
\label{sec:astro_rotation}

Lopsidedness decreases with $V_{\rm max}$ ($\rho = -0.39$, $p \approx 5\times10^{-34}$; Section \ref{sec:dynamical}), but at fixed stellar mass the dependence on $V_{\rm max}$ vanishes, so this is the mass trend rather than an independent dynamical one. Two explanations are commonly offered for such a trend. Faster rotation shortens the time over which an $m=1$ perturbation winds up and phase-mixes \citep{2014RvMP...86....1S}, and lower-mass discs
accrete gas at a higher rate relative to their stellar mass, so that misaligned accretion excites $m=1$ modes more easily \citep{2005A&A...438..507B}.  Both are consistent with the data; the catalogue alone does not distinguish them.

\subsection{The $Q_{\rm dyn}$--$\Amone$ mass-ratio--asymmetry link}
\label{sec:astro_stability}

We find a positive correlation between the dynamical mass ratio proxy $Q_{\rm dyn}$ (Eq. \ref{eq:Q_dyn}) and $\Amone$ ($\rho = +0.35$, $p \approx 10^{-27}$), which survives at fixed stellar mass with a reduced strength ($\rho_{\rm p} = +0.23$; $+0.18$ for $A_{1,\rm corr}$). Because $Q_{\rm dyn} = M_{\rm dyn}(<R_{V_{\rm max}})/M_\star$, this trend indicates that systems with relatively higher enclosed dynamical mass compared to their stellar content -- i.e.\ galaxies whose stellar component is a smaller fraction of the mass budget within $R_{V_{\rm max}}$, characteristic of dwarf and dark-matter-dominated discs -- tend to have the largest global $\Amone$ amplitudes. A natural interpretation is that a disc whose self-gravity is a small part of the potential responds more strongly to an offset or perturbed halo and to external perturbations (cold gas accretion, minor mergers, tidal interactions), and that its $m=1$ modes are less efficiently damped \citep{2009PhR...471...75J}. An alternative that is sometimes invoked, in which strong self-gravitating $m=2$ structures in massive discs suppress the $m=1$ signal \citep{1986MNRAS.221..213A,2014RvMP...86....1S}, is not supported by
our data: the median $\Amtwo$ is independent of both $V_{\rm max}$ ($\rho = -0.05$, $p = 0.10$; Fig. \ref{fig:fig9_rotation_analysis}b Fig. \ref{fig:fig9_rotation_analysis}) and stellar mass ($\rho = -0.01$), although the fraction of galaxies with a coherent $m=2$ pattern rises from 61 to 83 \% between $\log_{10}(M_\star/\Msun) = 9.5$--10 and 10.5--11.  The strength of $m=2$ patterns therefore appears to be set by internal disc dynamics rather than by the total dynamical mass, consistent with the slow secular evolution of bars once formed \citep{1981A&A....96..164C,1993RPPh...56..173S,2000ApJ...543..704D} and with the observed weak dependence of bar frequency on dynamical mass \citep{2018MNRAS.474.5372E}.  Bar populations in IllustrisTNG itself have been characterised by \citet{2020MNRAS.491.2547R}, and fast bars remain a point of tension with $\Lambda$CDM haloes \citep{2021A&A...650L..16F}.

\subsection{Mass dependence and cosmological context}
\label{sec:mass_dependence}

The systematic decline of median $\Amone$ with stellar mass -- from $\approx 0.115$ at $\log_{10}(M_\star/\Msun) \sim 9.7$ to $\approx 0.06$ at $\sim 10.7$ -- reproduces qualitatively the trend reported by \citet{2013ApJ...772..135Z} for their S$^{4}$G sample. In the cosmological context of \citet{2009PhR...471...75J}, lopsidedness is understood as a fossil record of the disc's accretion and interaction history. Minor mergers, tidal encounters, and cosmic-web gas accretion all imprint transient $m=1$ modes that decay on timescales of a few $t_{\rm dyn}$. The mass dependence we recover is consistent with the picture in which massive discs have completed more dynamical relaxation cycles and are closer to axisymmetric equilibrium, while low-mass systems still bear the marks of their most recent accretion or interaction event; as noted in Section \ref{sec:astro_rotation}, the catalogue alone does not distinguish this from a higher rate of perturbation in low-mass discs.

TNG50-1 is statistically consistent with the observed distribution of $\Amone$ at the mass-matched level (which, with 35 observed galaxies, does not mean the two distributions are identical), reproduces the observed decrease of $\Amone$ with stellar mass, and correlations with $V_{\rm max}$ and $Q_{\rm dyn}$ that are at least qualitatively expected.  The natural next step is to use the merger trees to connect individual $\Amone$ measurements to the recent accretion and interaction history of each galaxy, and so to measure how long specific channels keep a disc lopsided.

\section{Conclusions}
\label{sec:conclusions}

We have presented \pipeline, an open-source pipeline for the Fourier analysis of simulated disc galaxies, and applied it with identical settings to TNG50-1 and TNG100-1. Our main results are as follows.

\begin{enumerate}
  \item The pipeline recovers injected Fourier amplitudes in four synthetic galaxy families to better than 2 per cent at $N = 5\times10^5$ particles, with residuals following the expected $N^{-1/2}$ scaling (Section \ref{sec:validation}).

  \item Fourier amplitudes are biased upwards by shot noise, and the bias depends on resolution.  A per-galaxy noise floor, obtained by randomising particle azimuths, and the quadrature-corrected amplitude $A_{1,\rm corr}$ remove this bias: TNG50-1 discs downsampled to the TNG100-1 particle mass have uncorrected amplitudes 43--60 per cent too high, but their $A_{1,\rm corr}$ agrees with the full-resolution value to within 2--13 per cent (Sections \ref{sec:noise_floor} and \ref{sec:resolution}).  The bootstrap mean adds a further upward bias and should not be used as the amplitude estimate.

  \item For the 915 TNG50-1 galaxies with $3.16\times10^9 < M_\star/\Msun < 10^{11}$, the median bootstrap-mean $\Amone$ is 0.097, the median $A_1^{\rm lit}$ 0.091 and the median $A_{1,\rm corr}$ 0.086.  The distribution is consistent with the mass-matched S$^{4}$G sample of \citet{2013ApJ...772..135Z} ($D = 0.214$, $p = 0.079$), with the caveat that this sample contains only 35 galaxies.

  \item With the same pipeline, TNG100-1 is more lopsided than TNG50-1, not less. The TNG100/TNG50 ratio of medians falls from 2.31 for the bootstrap mean to 2.06 for $A_1^{\rm lit}$ and 1.59 for $A_{1,\rm corr}$ in the parent samples (1.64, 1.55 and 1.41 for kinematically selected discs). At fixed stellar mass the corrected excess is 1.5--1.9.

  \item The residual excess persists whether the aperture is defined in units of $R_{\rm d}$ or in kpc (1.5--1.7 at 5--8\,kpc, 1.8--2.2 at 3.5--6\,kpc), is not explained by the hotter kinematics of the TNG100-1 discs, and is largest at 3.5--6\,kpc.  It is consistent with an effect of the lower resolution of TNG100-1, for example its larger softening length or the discreteness noise of its heavier dark-matter particles, but differences between the two galaxy populations have not been excluded.

  \item The low median of \citet{2022A&A...662A..53L} for TNG100 cannot be attributed to a resolution deficit on the basis of a cross-study comparison, because aperture, centring, disc selection and possibly the mode normalisation differ. Our matched-pipeline analysis finds higher corrected amplitudes in TNG100-1 than in TNG50-1.

  \item In TNG50-1, $\Amone$ decreases with stellar mass and $V_{\rm max}$ (the latter through its correlation with mass) and increases with the dynamical-to-stellar mass ratio $Q_{\rm dyn}$ even at fixed mass; $\Amtwo$ shows no dependence on $V_{\rm max}$.
\end{enumerate}

These results have a practical consequence for any comparison of Fourier amplitudes across simulations or between simulations and observations: the estimator, the aperture and the particle number must be controlled, and positive-definite amplitudes should be corrected for their noise floor before they are compared.

Future work will (i) extend the comparison to TNG300-1, where the noise floor is larger still; (ii) apply the pipeline to other simulation suites such as EAGLE \citep{2015MNRAS.446..521S} and FIRE-2 \citep{2018MNRAS.480..800H}.

The \pipeline\ code and the TNG50-1 and TNG100-1 catalogues, including the noise-corrected amplitudes and the $D/T$ tables, are publicly available at
\url{https://github.com/Abbos998/TNGGalaxyLab} (Zenodo DOI: \href{https://doi.org/10.5281/zenodo.21363951}{10.5281/zenodo.21363951}).

\section*{Data Availability}

The TNG50-1 and TNG100-1 data used are publicly available via the IllustrisTNG data release~\citep{2019ComAC...6....2N} at \url{https://www.tng-project.org/data/}.  The \citet{2013ApJ...772..135Z} catalogue is available via VizieR (J/ApJ/772/135).  The processed catalogues supporting this paper
(the 915-galaxy TNG50-1 and 6717-galaxy TNG100-1 lopsidedness catalogues, including bootstrap, direct and noise-corrected amplitudes, the $D/T$ tables, and the inputs and scripts of the resolution tests of Section \ref{sec:resolution}), together with the full \pipeline\ source code, are available at
\url{https://github.com/Abbos998/TNGGalaxyLab} (Zenodo DOI \href{https://doi.org/10.5281/zenodo.21363951}{10.5281/zenodo.21363951}). The repository includes a \texttt{requirements.txt} for pip installation, an \texttt{environment.yml} for conda, a \texttt{Dockerfile} for container-based execution, and a continuous-integration workflow that automatically validates the pipeline on Linux, macOS, and Windows with Python 3.11 and 3.12.  All figures in this paper can be regenerated by running the validation and figure-generation scripts distributed in the repository; the results in this version correspond to release v17 (commit \texttt{9a6dab1}).

\section*{Acknowledgements}
A. U. Omonov  thanks the IllustrisTNG collaboration for providing public access to the simulation data. This research made extensive use of the NASA Astrophysics Data System (ADS). We made use of the IllustrisTNG public data release \citep{2019ComAC...6....2N} and the S$^{4}$G data release \citep{2010PASP..122.1397S,2015ApJS..217...32B}. The development of \pipeline\ was supported by the National University of Uzbekistan.

This work was supported by the National University of Uzbekistan, the Ministry of Higher Education, Science and Innovation of the Republic of Uzbekistan through Grant FZ-2020092851, and budgetary funding from the Academy of Sciences of the Republic of Uzbekistan.

We acknowledge use of the open-source software \textsc{NumPy} \citep{2020Natur.585..357H}, \textsc{SciPy} \citep{2020NaMet..17..261V}, \textsc{Matplotlib} \citep{2007CSE.....9...90H}, \textsc{Astropy} \citep{2022ApJ...935..167A}, and \textsc{H5PY}.

\bibliographystyle{mnras}
\bibliography{example}

\appendix

\section{Normalisation derivation}
\label{app:norm}

The factor of 2 in Eq. \ref{eq:Am_phim} arises as follows.  For a sinusoidal perturbation of amplitude $\varepsilon_m$, Eq. \ref{eq:fourier} gives
$\Sigma = \Sigma_0[1 + \varepsilon_m \cos(m\phi - m\phi_m)]$.  The continuous Fourier coefficient of this expansion is $c_m^{\rm continuous} = \varepsilon_m \Sigma_0 / 2$.  Since $c_0^{\rm continuous} = \Sigma_0$, the ratio $|c_m|/c_0 = \varepsilon_m / 2$.  Multiplying by 2 recovers the fractional amplitude $\varepsilon_m$, so that a $30\%$ perturbation reads out as $\Amone = 0.30$.

\section{Catalogue columns and excerpt}
\label{app:catalog_columns}

The complete 915-galaxy TNG50-1 catalogue is released as a machine-readable comma-separated file (\texttt{catalog.csv}, 29 columns) and as an equivalent FITS binary table (\texttt{catalog.fits}) via the code repository (see Data Availability).  Version 2 of the catalogue (\texttt{catalog\_v2.csv}, 45 columns), released for both TNG50-1 and the 6717-galaxy TNG100-1 sample, adds the noise-floor and noise-corrected columns of Section \ref{sec:noise_floor}; the original columns are unchanged.  Column definitions are given in Table \ref{tab:catalog_columns}; a truncated excerpt showing the twenty most lopsided galaxies of the TNG50-1 parent sample, sorted by bootstrap-mean $\Amone$, is given in Table \ref{tab:catalog_excerpt}.

\begin{table*}
  \centering
 \caption{Primary columns of the TNG50-1 and TNG100-1 catalogues. The original release (\texttt{catalog.csv}) contains 29 columns per galaxy and version 2 (\texttt{catalog\_v2.csv}) 45; the ones listed below are the fields referenced in this paper. Additional columns include integral-variant amplitudes ($A_m^{\rm integral,area}$, $A_m^{\rm integral,Jog}$), the $m=2$ phase-scatter diagnostic, bar-length bootstrap statistics, and, in version 2, the noise floors and noise-corrected amplitudes in the two fixed physical apertures.}
  \label{tab:catalog_columns}
  \begin{tabular}{lll}
    \toprule
    Column & Unit & Description \\
    \midrule
    \texttt{subhalo\_id}          & ---     & TNG50-1 subhalo ID at snapshot 99 \\
    \texttt{M\_star\_Msun}        & $\Msun$ & Total stellar mass of the subhalo \\
    \texttt{R\_d\_kpc}            & kpc     & Exponential disc scale length \\
    \texttt{aperture\_R\_in\_kpc}  & kpc     & Inner aperture radius ($1.5\,R_d$) \\
    \texttt{aperture\_R\_out\_kpc} & kpc     & Outer aperture radius ($2.5\,R_d$) \\
    \texttt{A1\_literature}       & ---     & $\langle\Amone\rangle$ R\&Z95 aperture average \\
    \texttt{A1\_bootstrap\_mean}  & ---     & Bootstrap-mean $\Amone$ \\
    \texttt{A1\_bootstrap\_std}   & ---     & Bootstrap 1$\sigma$ uncertainty \\
    \texttt{A2\_bootstrap\_mean}  & ---     & Bootstrap-mean $\Amtwo$ \\
    \texttt{A2\_bootstrap\_std}   & ---     & Bootstrap 1$\sigma$ uncertainty on $\Amtwo$ \\
    \texttt{bar\_length\_kpc}     & kpc     & Phase-coherent bar length (0 if none) \\
    \texttt{bar\_angle\_deg}      & deg     & Position angle of the bar \\
    \texttt{pattern\_coherence}   & ---     & $f_{\rm coh}$ quality flag ($m=2$) \\
    \texttt{v\_max\_kms}          & $\kms$  & Peak circular velocity (tracer method) \\
    \texttt{n\_disk\_particles}   & ---     & Star particles inside $|z|<3\kpc$, $R<30\kpc$ \\
    \midrule
    \multicolumn{3}{l}{\emph{Added in version 2 (\texttt{catalog\_v2.csv})}} \\
    \texttt{A1\_noise}           & ---     & Shot-noise floor $A_{1,\rm noise}$ in $[1.5,2.5]\,R_{\rm d}$ (Section~\ref{sec:noise_floor}) \\
    \texttt{A1\_corr}            & ---     & Noise-corrected $A_{1,\rm corr}$, Eq.~(\ref{eq:a1corr}) \\
    \texttt{A1\_bootstrap\_bc}  & ---     & Bias-corrected bootstrap estimate $2A_1^{\rm lit}-\langle\hat A_1\rangle_{\rm boot}$ \\
    \texttt{neff\_aperture}      & ---     & Effective particle number in the aperture \\
    \texttt{A1\_\{lit,noise,corr\}\_3.5\_6kpc} & --- & As above, fixed aperture 3.5--6\,kpc \\
    \texttt{A1\_\{lit,noise,corr\}\_5\_8kpc}   & --- & As above, fixed aperture 5--8\,kpc \\
    \texttt{pbc\_fixed}          & ---     & 1 if re-processed after periodic-boundary unwrapping \\
    \bottomrule
  \end{tabular}
\end{table*}

\begin{table*}
  \centering
  \caption{Twenty most lopsided galaxies of the TNG50-1 parent sample, sorted by bootstrap-mean $\Amone$ in descending order.  Bootstrap uncertainties are quoted at $1\sigma$ from $N_{\rm boot}=200$ resamples.  Following Section \ref{sec:bar_length}, the phase-coherent bar length $L_{\rm bar}$ is used only through the detection dichotomy $f_{\rm coh} > 0.5$ vs.\ $f_{\rm coh} \le 0.5$ and is not tabulated here; the full machine-readable catalogue reports it as a distinct column (\texttt{bar\_length\_kpc}, see Table \ref{tab:catalog_columns}). The complete 915-galaxy catalogue is available via the code repository (Zenodo DOI: 10.5281/zenodo.21363951).}
  \label{tab:catalog_excerpt}
  \begin{tabular}{rccccc}
    \toprule
    Subhalo ID  &  $M_\star$ [$\Msun$]  &  $\Rd$ [kpc]  &
    $\langle \Amone \rangle$  &  $\langle \Amtwo \rangle$  &
    $f_{\rm coh}$ \\
    \midrule
    668581  &  $5.37\!\times\!10^{9}$  &  1.73  &  $0.952 \pm 0.007$  &  $0.302 \pm 0.008$  &  0.91  \\
    661532  &  $3.24\!\times\!10^{9}$  &  2.04  &  $0.499 \pm 0.022$  &  $0.803 \pm 0.015$  &  1.00  \\
    646803  &  $3.97\!\times\!10^{9}$  &  1.57  &  $0.455 \pm 0.011$  &  $0.105 \pm 0.012$  &  0.55  \\
    558067  &  $2.03\!\times\!10^{10}$  &  2.60  &  $0.448 \pm 0.007$  &  $0.502 \pm 0.007$  &  0.81  \\
    650623  &  $4.87\!\times\!10^{9}$  &  2.67  &  $0.391 \pm 0.011$  &  $0.111 \pm 0.011$  &  0.69  \\
    620172  &  $4.29\!\times\!10^{9}$  &  2.58  &  $0.380 \pm 0.016$  &  $0.189 \pm 0.020$  &  0.89  \\
    655323  &  $6.02\!\times\!10^{9}$  &  2.42  &  $0.377 \pm 0.009$  &  $0.194 \pm 0.009$  &  0.91  \\
    645391  &  $5.79\!\times\!10^{9}$  &  3.82  &  $0.370 \pm 0.010$  &  $0.237 \pm 0.012$  &  0.56  \\
    657568  &  $4.36\!\times\!10^{9}$  &  2.70  &  $0.368 \pm 0.012$  &  $0.229 \pm 0.010$  &  0.86  \\
    647768  &  $4.52\!\times\!10^{9}$  &  2.64  &  $0.367 \pm 0.013$  &  $0.336 \pm 0.013$  &  0.73  \\
    663812  &  $6.39\!\times\!10^{9}$  &  2.01  &  $0.350 \pm 0.011$  &  $0.073 \pm 0.011$  &  0.44  \\
    594246  &  $6.38\!\times\!10^{9}$  &  1.84  &  $0.330 \pm 0.009$  &  $0.220 \pm 0.009$  &  0.41  \\
    663308  &  $5.06\!\times\!10^{9}$  &  1.82  &  $0.329 \pm 0.010$  &  $0.084 \pm 0.009$  &  0.53  \\
    661449  &  $5.21\!\times\!10^{9}$  &  4.18  &  $0.327 \pm 0.009$  &  $0.214 \pm 0.010$  &  0.38  \\
    578834  &  $1.64\!\times\!10^{10}$  &  3.75  &  $0.316 \pm 0.007$  &  $0.212 \pm 0.007$  &  0.73  \\
    600743  &  $9.61\!\times\!10^{9}$  &  2.59  &  $0.305 \pm 0.016$  &  $0.203 \pm 0.014$  &  0.93  \\
    663231  &  $3.28\!\times\!10^{9}$  &  2.47  &  $0.300 \pm 0.016$  &  $0.098 \pm 0.015$  &  0.88  \\
    715911  &  $5.04\!\times\!10^{9}$  &  1.35  &  $0.298 \pm 0.010$  &  $0.082 \pm 0.008$  &  0.62  \\
    670767  &  $4.13\!\times\!10^{9}$  &  1.90  &  $0.295 \pm 0.014$  &  $0.065 \pm 0.012$  &  0.27  \\
    638742  &  $7.03\!\times\!10^{9}$  &  2.22  &  $0.290 \pm 0.008$  &  $0.067 \pm 0.009$  &  0.57  \\
    \bottomrule
  \end{tabular}
\end{table*}

\section{Independent reproducibility audit}
\label{app:reproducibility_audit}

Table \ref{tab:reproducibility_audit} lists the principal statistics quoted in the main text alongside the values recomputed directly from the released catalogue (\texttt{catalog.csv}).  All computations were performed with an independent Python re-implementation using only the standard library
(\texttt{csv}, \texttt{statistics}, \texttt{math}), without recourse to any part of the \pipeline\ codebase.  Spearman rank correlations were evaluated with a small custom function identical in behaviour to \texttt{scipy.stats.spearmanr}.

\begin{table*}
  \centering
  \caption{Comparison of manuscript statistics with values independently recomputed from the released catalogue. Bootstrap ranks, correlations, and threshold-based sample counts reproduce the tabulated values to the precision printed in the main text.  Small differences in the fourth decimal place reflect rounding of the printed quantities; no discrepancy exceeds the corresponding bootstrap uncertainty.}
  \label{tab:reproducibility_audit}
  \begin{tabular}{lccc}
    \toprule
    Statistic & Value in paper & Recomputed & Difference \\
    \midrule
    Median $\Amone$                                      &     0.0970 &     0.0974 &     0.0004 \\
    Mean $\Amone$                                        &     0.1120 &     0.1117 &     0.0003 \\
    $1\sigma$ of $\Amone$ distribution                   &     0.0700 &     0.0702 &     0.0002 \\
    $16$th percentile of $\Amone$                        &     0.0560 &     0.0560 &     0.0000 \\
    $84$th percentile of $\Amone$                        &     0.1610 &     0.1606 &     0.0004 \\
    \midrule
    Galaxies with $\Amone > 0.1$                         &        431 &        431 &          0 \\
    Galaxies with $\Amone > 0.2$                         &         86 &         86 &          0 \\
    Galaxies with $f_{\rm coh} > 0.5$                    &        610 &        610 &          0 \\
    Overlap: $\Amone > 0.1$ \emph{and} $f_{\rm coh} > 0.5$ &        286 &        286 &          0 \\
    \midrule
    $\rho_{\rm Spearman}(Q_{\rm dyn}, \Amone)$           &     +0.350 &     +0.350 &      0.000 \\
    $\rho_{\rm Spearman}(V_{\rm max}, \Amone)$           &     -0.390 &     -0.387 &      0.003 \\
    $\rho_{\rm Spearman}(V_{\rm max}, \Amtwo)$           &     -0.050 &     -0.054 &      0.004 \\
    Median $Q_{\rm dyn}$                                 &       2.60 &       2.65 &       0.05 \\
    \bottomrule
  \end{tabular}
\end{table*}

The auditor's script is included in the code repository as \texttt{scripts/reproducibility\_audit.py} and can be re-executed by any reader with a copy of \texttt{catalog.csv} and a standard Python 3 interpreter.  We invite the community to extend this audit table with any further statistic of interest.

\section{Supplementary figures}
\label{app:supplementary}

\begin{figure*}
  \centering
  \includegraphics[width=\columnwidth]{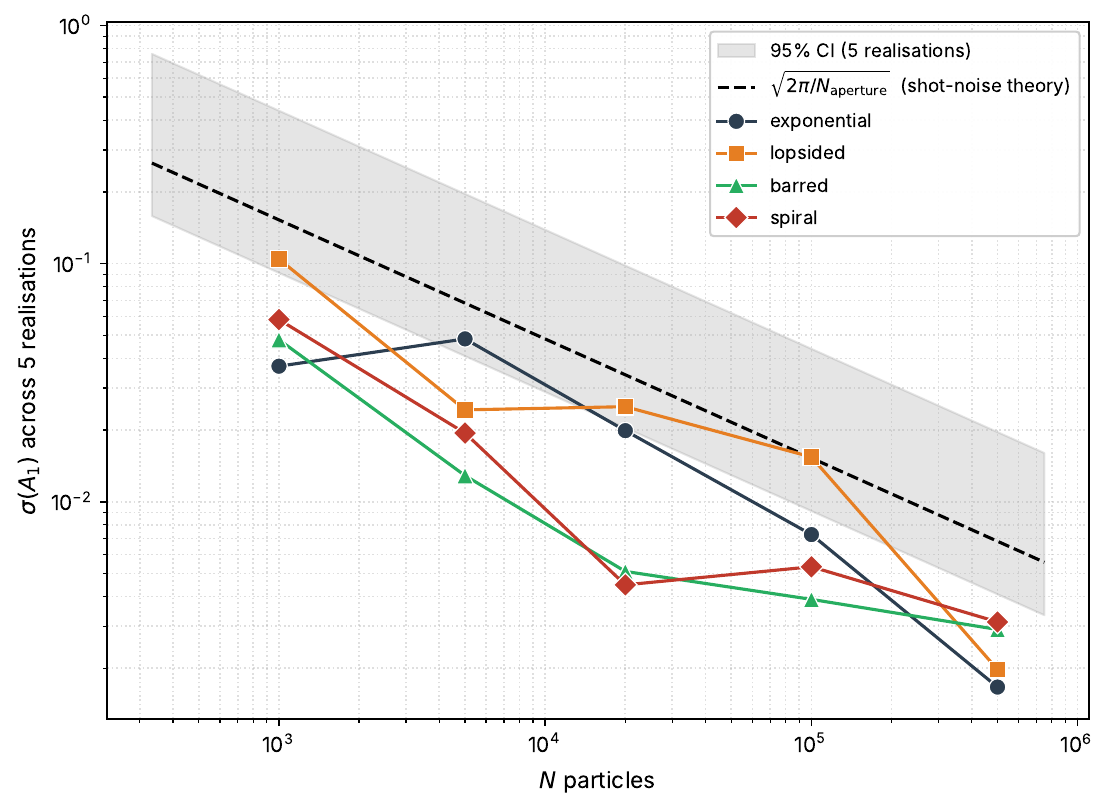}
  \caption{Empirical scatter $\sigma(\Amone)$ across five realisations vs.\ $N$ for the four synthetic families.  Dashed line: conservative reference scale $\sqrt{2\pi/N_{\rm aperture}}$ for the scatter (not to be confused with the mean noise amplitude $\simeq\sqrt{\pi/N}$ per annulus of Section \ref{sec:noise_floor}); empirical points lie at or below it because the aperture average combines several radial bins.  Grey band: 95\% CI for $\sigma$ estimated from five samples.}
  \label{fig:fig2_shot_noise_scaling}
\end{figure*}

\begin{figure*}
  \centering
  \includegraphics[width=\textwidth]{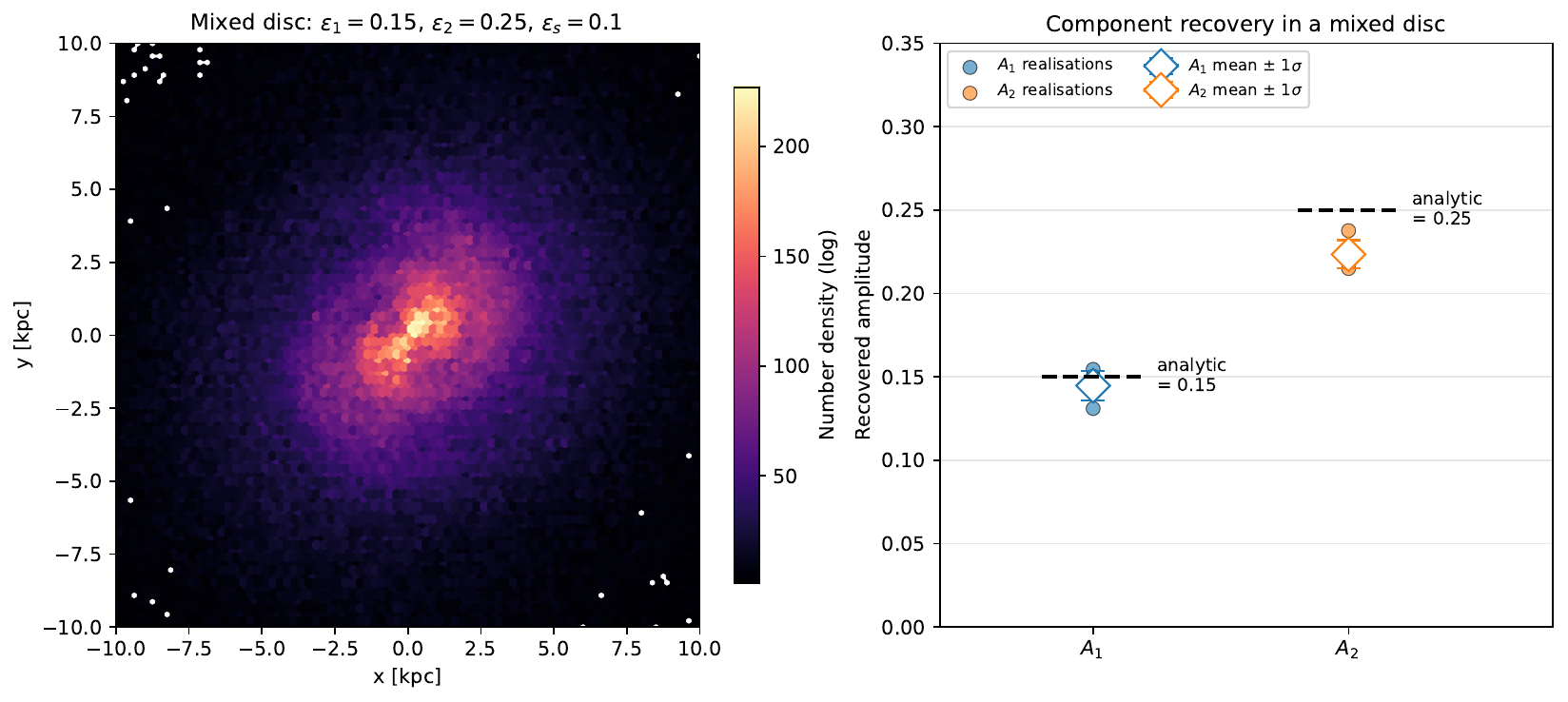}
  \caption{Mixed-component validation of the \pipeline\ pipeline. \textbf{Left}: surface-density map (log-scaled hexagonal binning) of a reference mixed disc with simultaneous $\varepsilon_1 = 0.15$ (lopsided), $\varepsilon_2 = 0.25$ (bar, oriented $45^{\circ}$ from the lopsided asymmetry), and $\varepsilon_s = 0.10$ (logarithmic spiral, pitch $15^{\circ}$), generated by rejection sampling with $N = 2\times 10^5$ particles for visualisation.  \textbf{Right}: recovered global amplitudes $\Amone$ (blue) and $\Amtwo$ (orange) across five independent realisations of the reference mixed disc at $N = 10^5$.  Small filled circles: individual realisations; open diamonds: mean $\pm 1\sigma$; dashed black ticks: analytically injected values.  The pipeline recovers each component to within $\sim 5\%$ despite the coexistence of the other modes.  The $-11\%$ bias in the recovered $\Amtwo$ reflects partial destructive interference from the injected logarithmic-spiral component ($\varepsilon_s = 0.1$, pitch angle $15^\circ$), which contributes to the $m=2$ mode with a radially varying phase; five random realisations are shown per configuration.}
  \label{fig:fig8_mixed_validation}
\end{figure*}

\begin{figure*}
  \centering
  \includegraphics[width=\textwidth]{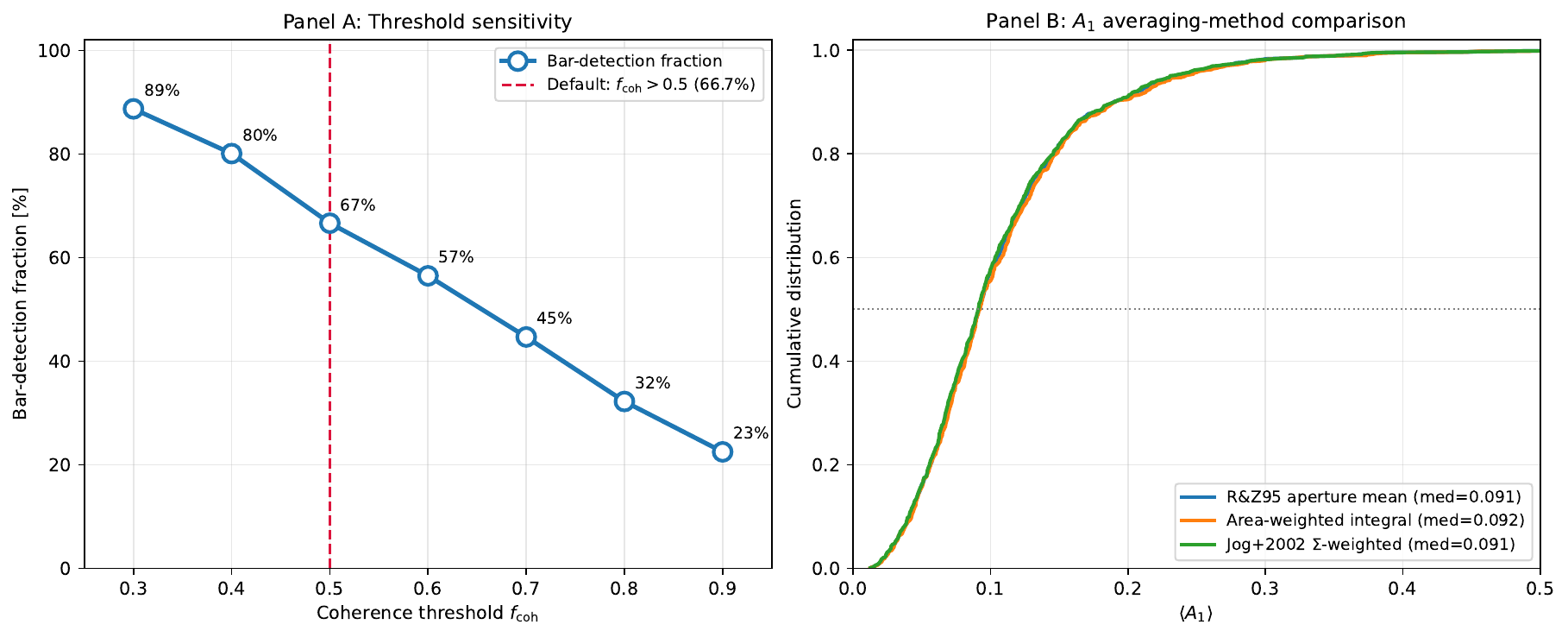}
  \caption{Parameter-sensitivity analyses on the 915-galaxy TNG50 catalogue.  \textbf{Panel A}: fraction of galaxies flagged as coherent-$m=2$ detections as a function of the coherence threshold $f_{\rm coh}^{\rm thresh}$; dashed red line marks the fiducial choice $f_{\rm coh}^{\rm thresh} = 0.5$ ($66.7\%$). \textbf{Panel B}: cumulative distribution of $\Amone$ under three amplitude-averaging conventions --- R\&Z95 aperture mean (blue), area-weighted integral (orange), and Jog+2002 $\Sigma$-weighted integral (green).  All three converge to within a few per cent across the full amplitude range. The Panel B medians ($0.091$--$0.092$) are computed on raw aperture-averaged amplitudes without bootstrap resampling, and therefore differ at the few-per-cent level from the bootstrap-mean median $0.097$ quoted in Section \ref{sec:amplitude_distribution}.}
  \label{fig:fig10_parameter_sensitivity}
\end{figure*}

\begin{figure*}
  \centering
  \includegraphics[width=\textwidth]{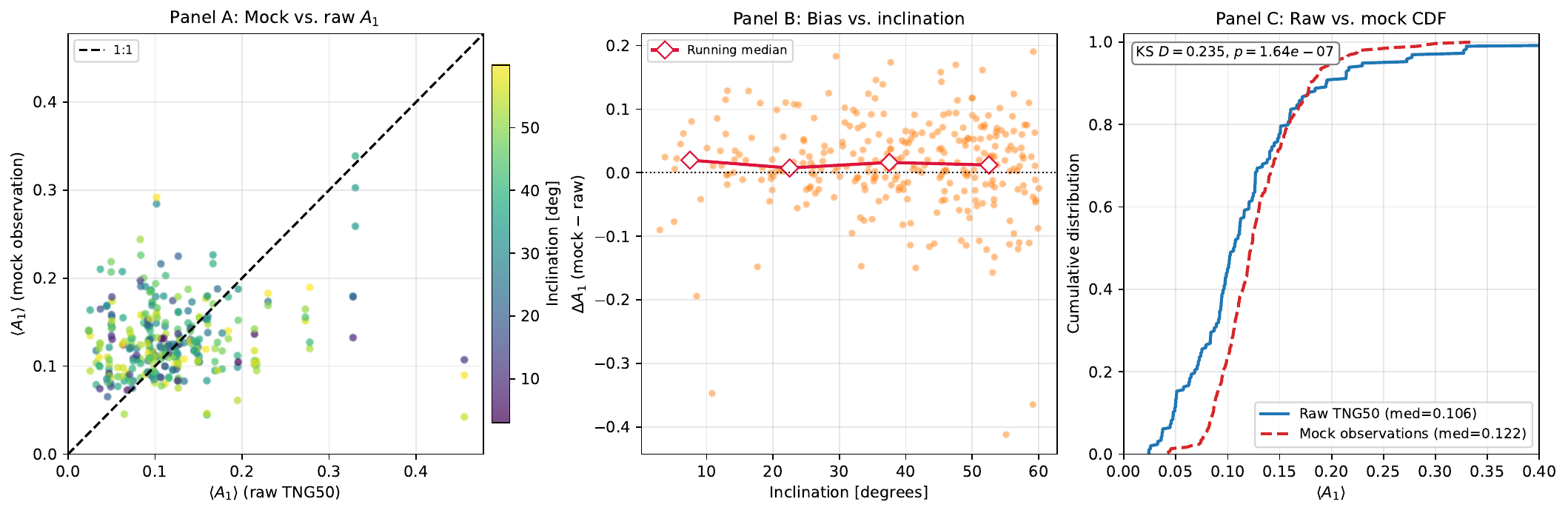}
  \caption{Mock-observation comparison of a 100-galaxy TNG50 subsample. Each galaxy is observed under three random (inclination, distance) combinations to generate an S$^{4}$G-like mock analogue.  \textbf{Panel A}: recovered $\Amone$ in the mock vs.\ the raw TNG50 measurement, colour-coded by inclination angle.  \textbf{Panel B}: bias $\Amone^{\rm mock} - \Amone^{\rm raw}$ as a function of inclination; red diamonds show the running median.  \textbf{Panel C}: cumulative distributions of $\Amone$ in the raw TNG50 sample (blue solid) and the mock-observed subsample (red dashed).  The raw median ($0.106$) refers to the specific 100-galaxy subsample used here and differs slightly from the full-sample median ($0.097$; see Section \ref{sec:amplitude_distribution}) due to random sampling.  Six mocks with $\Amone > 1$, which is not a physical amplitude, were excluded: they are all three realisations of two compact discs (subhaloes 595994 and 603004, $R_{\rm d} = 1.3$--$1.4\kpc$), for which the degraded image does not resolve the aperture.  Including them changes the mock median from 0.122 to 0.124.}
  \label{fig:fig11_mock_comparison}
\end{figure*}

Figures \ref{fig:fig2_shot_noise_scaling}--\ref{fig:fig11_mock_comparison} support the validation and robustness tests described in Sections \ref{sec:validation} and \ref{sec:mock_observations}.

\bsp
\label{lastpage}
\end{document}